\documentclass[sigconf, nonacm]{acmart}

\usepackage{amsmath,amsfonts}
\usepackage{textcomp}
\usepackage{xcolor}
\usepackage{xspace}
\usepackage{multirow}
\usepackage{bm}
\usepackage{subcaption}
\usepackage{enumitem}
\usepackage{algorithm}
\usepackage{algorithmic}
\usepackage{booktabs}
\usepackage{threeparttable}

\newcommand{\eat}[1]{}
\newcommand{\stitle}[1]{\vspace{1ex} \noindent{\bf #1}}

\newcommand{\kw}[1]{{\ensuremath {\mathsf{#1}}}\xspace}

\newcommand{\CS}{\kw{CS}}
\newcommand{\ACS}{\kw{ACS}}

\newcommand{\ACQ}{\kw{ACQ}}
\newcommand{\ATC}{\kw{ATC}}
\newcommand{\CTC}{\kw{CTC}}
\newcommand{\ECC}{\kw{ECC}}
\newcommand{\AGG}{\kw{AGG}}
\newtheorem*{example}{Example}

\newcommand{\SQD}{\kw{Simple} \kw{QD}-\kw{GNN}}
\newcommand{\QD}{\kw{QD}-\kw{GNN}}
\newcommand{\AQD}{\kw{AQD}-\kw{GNN}}

\newcommand{\blue}[1]{{\color{black}{#1}}}

\newcommand\vldbdoi{10.14778/3514061.3514070}
\newcommand\vldbpages{XXX-XXX}
\newcommand\vldbvolume{15}
\newcommand\vldbissue{6}
\newcommand\vldbyear{2022}
\newcommand\vldbtitle{\shorttitle} 
\newcommand\vldbavailabilityurl{}
\newcommand\vldbpagestyle{empty}

\begin{document}
% \title{Graph Neural Networks for Attributed Community Search}
% \title{Attributed Community Search with Attributed Query Driven Graph Neural Networks}
\title{Query Driven-Graph Neural Networks for Community Search: From Non-Attributed, Attributed, to Interactive Attributed}

%%
%% The "author" command and its associated commands are used to define the authors and their affiliations.

% \author[*1]{Yuli Jiang}
% \author[$\dag$2]{Yu Rong}
% \author[*3]{Hong Cheng}
% \author[$\ddag$4]{Xin Huang}
% \author[*5]{Kangfei Zhao}
% \author[$\dag$6]{Junzhou Huang}
% \affil[*]{ The Chinese University of Hong Kong, Hong Kong, China}
% \affil[$\dag$]{ Tencent AI Lab, Shen Zhen, China}
% \affil[$\ddag$]{ Hong Kong Baptist University, Hong Kong, China}
% \affil[$\ $]{\texttt{\footnotesize \{$^1${yljiang},$^3$hcheng,$^5$kfzhao\}@se.cuhk.edu.hk,$^2$yu.rong@hotmail.com,$^4$xinhuang@comp.hkbu.edu.hk,$^6$jzhuang@uta.edu}}
% \vspace{-0.2cm}
\author{
Yuli Jiang$^{*1}$,
Yu Rong$^{\dagger 2}$,
Hong Cheng$^{*3}$,
Xin Huang$^{\ddagger 4}$, 
Kangfei Zhao$^{\dagger 5}$,
Junzhou Huang$^{\dagger 6}$ }
\affiliation{
 \institution{$^*$The Chinese University of Hong Kong, $^{\dagger}$Tencent AI Lab, $^{\ddagger}$Hong Kong Baptist University, China } 
\city{\small \{$^1$yljiang,$^3$hcheng,$^5$kfzhao\}@se.cuhk.edu.hk, $^2$yu.rong@hotmail.com, $^4$xinhuang@comp.hkbu.edu.hk, $^5$zkf1105@gmail.com, $^6$jzhuang@uta.edu}
}

\begin{abstract}

Given one or more query vertices, Community Search (CS) aims to find densely intra-connected and loosely inter-connected structures containing query vertices. 
Attributed Community Search (ACS), a related problem, is more challenging since it finds communities with both cohesive structures and homogeneous vertex attributes. 
However, most methods for the CS task rely on inflexible pre-defined structures and studies for ACS treat each attribute independently. 
Moreover, the most popular ACS strategies decompose ACS into two separate sub-problems, i.e., the CS task and subsequent attribute filtering task. However, in real-world graphs, the community structure and the vertex attributes are closely correlated to each other. This correlation is vital for the ACS problem. In this vein, we argue that the separation strategy cannot fully capture the correlation between structure and attributes simultaneously and it would compromise the final performance.

In this paper, we propose Graph Neural Network (GNN) models for both CS and ACS problems, i.e., Query Driven-GNN (\QD) and Attributed Query Driven-GNN (\AQD). 
In \QD, we combine the local query-dependent structure and global graph embedding. In order to extend \QD to handle attributes, we model vertex attributes as a bipartite graph and capture the relation between attributes by constructing GNNs on this bipartite graph. With a Feature Fusion operator, \AQD processes the structure and attribute simultaneously and predicts communities according to each attributed query. 
Experiments on real-world graphs with ground-truth communities demonstrate that the proposed models outperform existing CS and ACS algorithms in terms of both efficiency and effectiveness. 
More recently, an interactive setting for CS is proposed that allows users to adjust the predicted communities.
We further verify our approaches under the interactive setting and extend to the attributed context. Our method achieves $2.37\%$ and $6.29\%$ improvements in F1-score than the state-of-the-art model without attributes and with attributes respectively.

\end{abstract}

\maketitle

\vspace{-0.3cm}
%%% do not modify the following VLDB block %%
%%% VLDB block start %%%
\pagestyle{\vldbpagestyle}
\begingroup\small\noindent\raggedright\textbf{PVLDB Reference Format:}\\
% \vldbauthors. \vldbtitle. PVLDB, \vldbvolume(\vldbissue): \vldbpages, \vldbyear.\\
Yuli Jiang, Yu Rong, Hong Cheng, Xin Huang,  Kangfei Zhao, Junzhou Huang. \vldbtitle. PVLDB, \vldbvolume(\vldbissue): \vldbpages, \vldbyear.\\
\href{https://doi.org/\vldbdoi}{doi:\vldbdoi}
\endgroup
\begingroup
\renewcommand\thefootnote{}\footnote{\noindent
This work is licensed under the Creative Commons BY-NC-ND 4.0 International License. Visit \url{https://creativecommons.org/licenses/by-nc-nd/4.0/} to view a copy of this license. For any use beyond those covered by this license, obtain permission by emailing \href{mailto:info@vldb.org}{info@vldb.org}. Copyright is held by the owner/author(s). Publication rights licensed to the VLDB Endowment. \\
\raggedright Proceedings of the VLDB Endowment, Vol. \vldbvolume, No. \vldbissue\ %
ISSN 2150-8097. \\
\href{https://doi.org/\vldbdoi}{doi:\vldbdoi} \\
}\addtocounter{footnote}{-1}\endgroup
%%% VLDB block end %%%

% % \vspace{-0.1cm}
%%% do not modify the following VLDB block %%
%%% VLDB block start %%%
\ifdefempty{\vldbavailabilityurl}{}{
\begingroup\small\noindent\raggedright\textbf{PVLDB Artifact Availability:}\\
\vspace{-0.1cm}
The source code and data %, and/or other artifacts 
have been made available at \url{https://github.com/lizJyl/Codes-for-Peer-Review-of-VLDB-August-337}.
\endgroup
}
%\vspace{-0.3cm}
%%% VLDB block end %%%

\section{Introduction}\label{sec.intro}
Graph is an essential data structure to represent entities and their relationships, e.g., social networks, protein-protein interaction networks, web graphs, and knowledge graphs, to name a few. 
Community, a subgraph of densely intra-connected and loosely inter-connected structure, naturally exists as a functional module in real-world graphs.
Community Search (CS) \cite{CS_survey,CS_BOOK/2019Huang,sozio2010community,cui2013online,huang2014querying,hu2016querying} is a vital application in graph analytics. Concretely,  given any query vertices, CS aims to find a vertex set with cohesive structure according to the query, i.e., query-dependent communities. 
Attributed Community Search (ACS), a related but more challenging problem, has attracted a lot of attention recently \cite{CS_survey,CS_BOOK/2019Huang,huang2017community,fang2016effective,huang2017attribute}. Given any query vertex and attribute set, ACS aims at finding query-dependent communities with homogeneous attributes, which means the community members share similar attributes with the query attributes.

%Moreover, current studies also suffer from two serious limitations of \emph{structure inflexibility} and \emph{attribute irrelevance}.
For the CS and ACS problems, existing studies suffer from two serious limitations, that is,  \textbf{\emph{structure inflexibility}} and \textbf{\emph{attribute irrelevance}}. 
Structure inflexibility refers to the problem that most community search models are based on a pre-defined subgraph pattern, such as $k$-core \cite{sozio2010community, cui2014local,fang2016effective}, $k$-truss \cite{huang2014querying, huang2017attribute, akbas2017truss}, $k$-clique \cite{cui2013online, yuan2017index}, and $k$-edge connected component (\ECC) \cite{chang2015index, hu2016querying}. The pre-defined subgraph pattern imposes a very rigid requirement on topological structure of communities, which may not perfectly hold in real-world communities.
%For example, in a Twitter network, the communities of a famous person and an ordinary person can be dramatically different in terms of the community size and topological structure. 
Attribute irrelevance means existing models treat each attribute independently \cite{fang2016effective,huang2017attribute}. However, in real graphs, vertex attributes are not independent of each other. Ignoring such implicit relations would harm the quality of queried communities.

\begin{figure}
    \centering
     \includegraphics[width=1.0\linewidth]{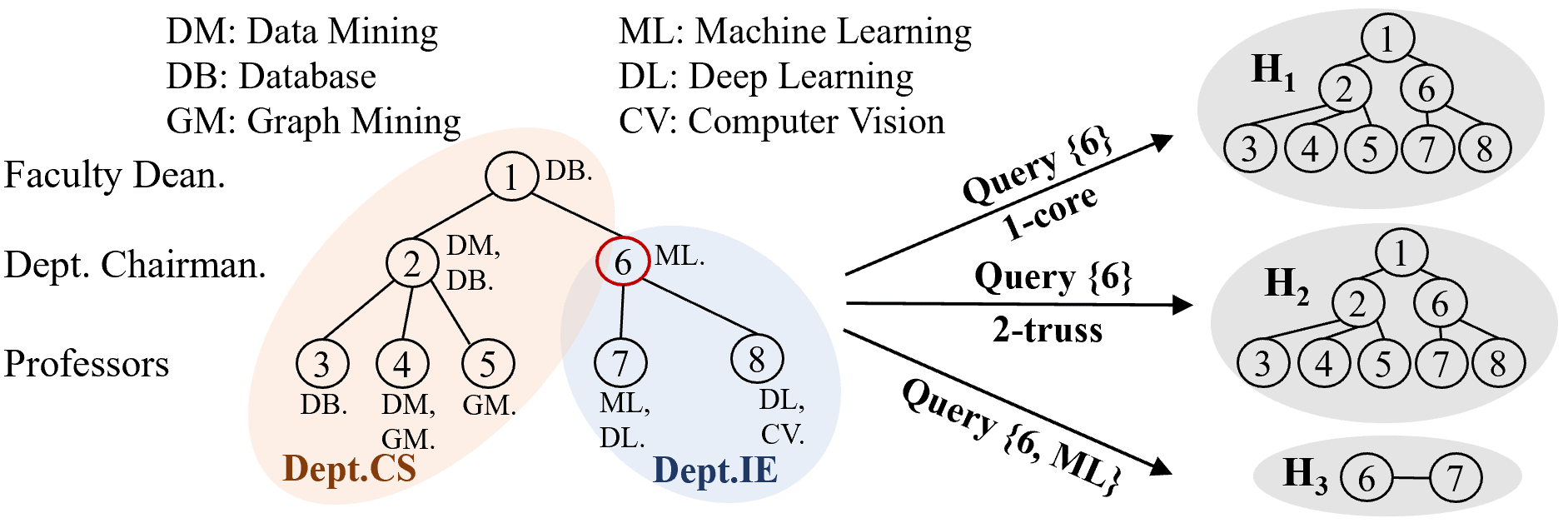}
	 \vspace{-0.8CM}
     \caption{An attributed graph depicting a faculty hierarchy with two departments: Dept.CS and Dept.IE. Attributes represent research topics. 
     For vertex $6$, there is a ground-truth Dept.IE community (shown in blue) as vertices $6-8$ are close and work on similar topics. 
     On the right, in response to queries on each arrow, there are three result communities found by existing algorithms, which are quite different from the ground-truth community.}%
     \label{fig:example1}
	 \vspace{-0.5CM}
    %  \vspace{-0.2CM}
\end{figure}

% Example. 
Figure \ref{fig:example1} depicts a toy example illustrating the limitation of existing algorithms. 
The faculty hierarchy is a tree-like structure from the faculty dean, department chairman to the professors in each department. \blue{Using existing methods based on pre-defined subgraph patterns, we can only find a $1$-core community of vertex $6$ in $H_1$ and a $2$-truss community in $H_2$, which are the entire graph. These $k$-core \cite{sozio2010community} and $k$-truss \cite{huang2014querying} patterns cannot discover the tree-like department communities owing to the structure inflexibility. 
For attributed community search, when querying the community of vertex 6 and attribute ``ML'', current methods \cite{fang2016effective,huang2017attribute} find the community $H_3$ since they ignore the implicit relations between ``ML'', ``DL'' and ``CV''.}  Thus, existing studies suffer from these two inadequacies on structure and attribute respectively.

% 2 -stage  separate  structure and attribute 
% not good
% 
%There exist several models \cite{fang2016effective,huang2017attribute} for attributed community search in the literature. 
%These methods take a two-stage process. 
%The existing studies \cite{fang2016effective,huang2017attribute} for ACS usually adopt a two-stage strategy which deals with the query vertices and attributes separately. 

%which deals with the query vertices and attributes separately. Specifically, 
%First, it finds a candidate community for query vertices with dense structure; Second, it shrinks the community by optimizing an attribute function. However, the two-stage strategy \emph{treats the cohesive structure and attribute homogeneity in a community separately} and ignores the correlation between structure and attributes. This may lead to inferior community search performance. For instance, in protein-protein interaction networks, proteins with similar functions (i.e., attributes) are more likely to interact with each other \cite{szklarczyk2015string}. We should not process them independently.

% community model not flexible

Moreover, for the ACS problem, existing studies \cite{fang2016effective,huang2017attribute} usually adopt a two-stage strategy which first finds the candidate community by considering the topological structure only, and then performs a filtering on the candidate community by considering the attribute similarity.  The two-stage strategy \emph{treats the structure cohesiveness and attribute homogeneity separately}.  But there is usually a correlation between structure and attribute, for instance, in protein-protein interaction networks, proteins with similar functions (i.e., attributes) are more likely to interact with each other \cite{szklarczyk2015string}.  Independently dealing with the structure and attribute would harm the quality of queried communities.

Inspired by the success of Graph Neural Network (GNN) \cite{Kipf_GCN} on combining attribute and structure in many graph problems, %we utilize GCN to solve this attributed community search problem. 
Gao \emph{et al.} \cite{gao2021ics_ICS-GNN} proposed a GNN-based framework, ICS-GNN, to solve the community search problem in an interactive fashion (i.e., users can adjust predicted communities during the query process). 
Specifically, it enhances the non-attributed queries by the GNN model \cite{Kipf_GCN} which exploits the information from the existing vertex attributes in graphs. However, for every query, ICS-GNN re-trains the whole model. This re-training process is time-consuming and hinders its applications in real-world scenarios, especially for the online query case. On the other hand, even though ICS-GNN makes use of the attributes to enhance the community search performance, its model architecture cannot accept the query attributes as input.  Therefore, ICS-GNN cannot be extended to support interactive attributed community search easily.

To address the above limitations, in this paper, we propose GNN-based models for both CS and ACS problems.  For the CS problem, to address the structure inflexibility issue,
we design a two-branch model: Query Driven-GNN (\QD) to encode the information from both the query and graph.  Concretely, \QD contains two encoders, \emph{Query Encoder} and \emph{Graph Encoder}. \emph{Query Encoder} encodes the structural information from query vertices and focuses on modeling the local topology around the queries. \emph{Graph Encoder} combines the global structure and attributes to learn the query-independent node embeddings. As a learning-based model, \QD can search communities without imposing any restriction on the community structure.  Furthermore, we design an additional \emph{Attribute Encoder} to extend \QD to support the attributed community search. \emph{Attribute Encoder} exploits a node-attribute bipartite graph to model the attribute relations and can encode more meaningful information from the attribute space. To process structure and attribute simultaneously, we employ a \emph{Feature Fusion} component to fuse the information from different encoders and make the final output. Furthermore, we design a new query framework which detaches the model training from the online query stage.  Therefore, our framework does not need the time-consuming re-training phase for online query applications.

To summarize, we make the following contributions.
\begin{itemize}[leftmargin=*]

    % \item We study the problem of attributed community search, that is, finding query-dependent communities with cohesive structure and homogeneous attributes w.r.t.\ the query vertices and query attributes.
	\vspace{-0.15CM}
    \item We propose a Query Driven-GNN model (\QD) for community search, which combines the local query-dependent structure and global node embeddings. Given any query, \QD only needs a model inference step and avoids the time-consuming re-training. 
    
    %\QD can acquire query information directly and process a group of queries at once.
    % predict new queries according to history.
    \vspace{-0.05CM}
    \item To the best of our knowledge, this is the first work that proposes a GNN model for the attributed community search problem, called Attributed Query Driven-GNN (\AQD). Our novel learning framework extends GNN into ACS through a node-attribute bipartite graph, and learns the community information from both the local structure and similar attributes of queries. 
    % Our novel supervised learning model is not built upon pre-defined dense subgraph patterns and fixed attribute optimizations, but learns community information from both the global graph and local 
    
    \vspace{-0.05CM}
    \item We conduct extensive experiments on real-world data sets with ground-truth communities for performance evaluation.
    % to evaluate our proposed model. 
    Experiments demonstrate that our model significantly outperforms state-of-the-art methods in terms of community quality with only 4.31 milliseconds average response time.% and query time. 
    
    \vspace{-0.05CM}
    \item We apply \AQD to the interactive community search problem and extend it into the attributed context. Experiments show that our models can improve the performance of ICS-GNN \cite{gao2021ics_ICS-GNN} in both non-attributed and attributed manner with $2.37\%$ and $6.29\%$ improvements in F1-score respectively.
    % \item We verify \QD and \AQD into the interactive setting of CS and ACS. We replace the learning model in ICS-GNN by our models, and experiments show that the proposed models can still improve the performances ICS-GNN in both non-attributed and attributed manner with $2.37\%$ and $6.29\%$ improvements in F1-score respectively.
    %We conduct extensive experiments on both real dataset and produced dataset. And experiments show that our method performance exceed all other attribute community search methods.
    \vspace{-0.1CM}
\end{itemize}
\vspace{-0.2CM}

\stitle{Roadmap}. The rest of the paper is organized as follows. 
Section \ref{sec.related} discusses related work. 
Section \ref{sec.pre} gives some preliminaries. 
Section \ref{sec.frame} presents the common framework of the proposed models. 
Section \ref{sec.CS-GCN} introduces the \QD model for community search problem, and Section \ref{sec.ACS-GCN} describes the \AQD model for attributed community search. 
We present the experimental results in Section \ref{sec.exp} and conclude the paper in Section \ref{sec.con}.

\vspace{-0.2cm}
\section{Related Work}\label{sec.related}
%The studies related to this work can be classified into community search, community detection, and graph neural network.
\vspace{-0.05cm}
Our study is closely related to community search (CS) and graph neural network (GNN).

\vspace{-0.15cm}
\stitle{Community Search}. The problem of CS \cite{sozio2010community} is to find densely connected communities containing the query vertices. A comprehensive survey of CS models and existing approaches can be found in \cite{CS_survey,CS_BOOK/2019Huang}.
Various community models have been proposed based on different cohesive graph patterns, including $k$-core \cite{sozio2010community, cui2014local}, $k$-truss \cite{huang2014querying, huang2015approximate, akbas2017truss}, $k$-clique \cite{cui2013online, yuan2017index}, and $k$-edge connected component (\ECC) \cite{chang2015index, hu2016querying}. These pre-defined cohesive metrics are inflexible and can be too loose (e.g., $k$-core) or too tight (e.g., $k$-clique) to capture the topology structure of communities.  If the real-world communities do not follow any of the above graph patterns, these models would fail to discover the true communities. A learning-based model ICS-GNN \cite{gao2021ics_ICS-GNN} has recently been proposed for interactive community search. ICS-GNN first finds a candidate subgraph starting from query vertices, then learns the node embeddings through applying GNN model on subgraph, and finally employs a BFS based algorithm to select the $k$-sized community with maximum GNN scores. ICS-GNN does not support attributed community search as the query only involves vertices but no attributes. It also needs to re-train the entire model for each query, which is costly for this online query problem.

For attributed community search, \ACQ \cite{fang2016effective} and \ATC \cite{huang2017attribute} have been proposed, which aim to discover communities that contain query vertices and have similar attributes to the query attributes. \ACQ is based on $k$-core and finds communities with the maximum number of common query attributes shared by community members. \ATC finds $k$-truss communities with the maximum pre-defined attribute score. Both adopt a two-stage process. They first impose a pre-defined structural constraint to find candidate communities, then optimize functions of attribute score to select the most related communities.  However, the attribute score functions ignore the similarities between attributes, and these two-stage methods fail to capture the correlation between structure and attribute. In this paper, we propose \QD, which considers the cohesive structure and homogeneous attributes in an integrated way. 

\vspace{-0.15cm}
\stitle{Graph Neural Network}. 
Inspired by the huge success of neural networks in natural language processing and computer vision, many graph analytic problems have been solved via graph neural networks \cite{Kipf_GCN}, such as node classification\cite{rong2020dropedge,chang2021spectral,he2019bipartite}, graph classification \cite{lijia_graphCliasf, NIPS2018_7707}, drug discovery \cite{ma2022cross, rong2020self,yu2021graph}, adversarial attacks \cite{BojchevskiG19, ZhuZ0019,chang2020restricted,chang2021not} and graph algorithmic tasks \cite{zhao2021learned, zhao2021finding}. To build good models, the advanced techniques of pooling \cite{LeeLK19, GaoJ19, 0001WAT19} and attention \cite{self_feat, LeeLK19, Wang00LC19} have been developed. However, most learning models are designed for specific tasks based on graph embedding \cite{LiSHZ18,Ye0FZW19} or end-to-end solutions \cite{ShangTHBHZ19, goel2019end}. 
Existing GNN models cannot extend to attributed community search straightforwardly. 
To the best of our knowledge, we are the first to propose a GNN-based model for attributed community search and extend ICS-GNN to the attributed context as well. 

% \vspace{-0.4cm}
\vspace{-0.2cm}
\section{Preliminaries}\label{sec.pre}
% In the following, we present the problem definition of community search. 
\vspace{-0.1cm}
In this section, we first introduce the notations and define the problems of CS and ACS formally, and then describe a general GNN as the foundation of our proposed models.

\vspace{-0.3cm}
\subsection{Definitions}
\label{Sec.defin}

\vspace{-0.05cm}
Let $G(\mathcal{V}, \mathcal{E})$ be a graph with a set $\mathcal{V}$ of vertices and a set $\mathcal{E} \subseteq \mathcal{V}\times \mathcal{V}$ of edges. Let $n=|\mathcal{V}|$ and $m=|\mathcal{E}|$ be the number of vertices and edges respectively. We denote $\mathcal{N}(\mathrm{v})=\{\mathrm{u} \text{ | } (\mathrm{u}, \mathrm{v})\in \mathcal{E}\}$ as the neighborhood set of vertex $\mathrm{v}$. Moreover, let $\mathcal{N}^{+}(\mathrm{v}) =\{\mathrm{v}\}\cup \mathcal{N}(\mathrm{v})$ be the vertex set containing $\mathrm{v}$'s neighbors and $\mathrm{v}$ itself.  %The adjacency matrix of $G$ is denoted as $\boldsymbol{A} \in \{0,1\}^{n\times n}$. For any pair of vertices $\mathrm{u}_i, \mathrm{u}_j\in V$, if there exists an edge $(\mathrm{u}_i, \mathrm{u}_j)\in \mathcal{E}$, $\boldsymbol{A}_{ij}=1$; otherwise, $\boldsymbol{A}_{ij}=0$. 

\vspace{-0.1cm}
\stitle{Community Search (CS)}. For a graph $G(\mathcal{V}, \mathcal{E})$, given a vertex query set $ \mathcal{V}_q \subseteq \mathcal{V}$, the problem of {Community Search}~(\CS) is to find the query-dependent community $\mathcal{C}_q \subseteq \mathcal{V}$. Vertices in community $\mathcal{C}_q$ need to be densely intra-connected, i.e., having cohesive structure.

Let $G(\mathcal{V}, \mathcal{E},\mathcal{F})$ be an attributed graph where $\mathcal{F}=\{ \mathcal{F}_1, \dots , \mathcal{F}_n\}$ is the set of vertex attributes and $\mathcal{F}_i$ is the attribute set of vertex $\mathrm{v}_i$. Define $\hat{\mathcal{F}}$ as the union of all the vertex attribute sets, i.e., $\hat{\mathcal{F}}= \mathcal{F}_1 \cup ... \cup \mathcal{F}_n$. Let $d$ be the number of unique attributes $d=|\hat{\mathcal{F}}|$.  The attribute set of each vertex, e.g., $\mathcal{F}_i$, is encoded to a $d$-dimensional vector $\boldsymbol{f}_i$. For a keyword attribute $\mathrm{f}_k$, if vertex $\mathrm{v}_i$ has this keyword, i.e., $\mathrm{f}_k \in \mathcal{F}_i$, then ${\boldsymbol{f}_i}_k=1$; otherwise, ${\boldsymbol{f}_i}_k=0$. For a numerical attribute $\mathrm{f}_j$, ${\boldsymbol{f}_i}_j$ is the value of vertex $\mathrm{v}_i$ on this attribute. Then the set of vertex attributes $\mathcal{F}=\{ \mathcal{F}_1, \dots , \mathcal{F}_n\}$ is encoded to an attribute matrix $\boldsymbol{F}=[\boldsymbol{f}_1, \dots, \boldsymbol{f}_n]^T \in \mathbb{R}^{n \times d}$.

\vspace{-0.1cm}
\stitle{Attributed Community Search (ACS)}. For an attributed graph $G(\mathcal{V}, \mathcal{E},$ $\mathcal{F})$, given a query $\langle \mathcal{V}_q,\ \mathcal{F}_q \rangle$ where $\mathcal{V}_q\subseteq \mathcal{V}$ is a set of query vertices, and $\mathcal{F}_q \subseteq \hat{\mathcal{F}}$ is a set of query attributes, the problem of {Attributed Community Search}~(\ACS) is to find the query-dependent community $\mathcal{C}_q \subseteq \mathcal{V}$. Vertices in community $\mathcal{C}_q$ need to be both structure cohesive and attribute homogeneous, i.e., vertices in a community are densely intra-connected in structure and attributes of these vertices are similar.

In this paper, we formulate the above two problems as a binary classification task. Given a query $q=\langle \mathcal{V}_q\rangle$ or $q=\langle \mathcal{V}_q,\ \mathcal{F}_q \rangle$, we classify the graph vertices into two classes (belonging to a community $\mathcal{C}_q$ of query $q$ or not).  We use the one-hot vector $\boldsymbol{c}_q\in \{0,1\}^{n}$ to represent the output community $\mathcal{C}_q$ by a model $\mathcal{M}$. If the output value $\boldsymbol{c}_{q_k}=1$, vertex $\mathrm{v}_k$ belongs to the result community $\mathcal{C}_q$ predicted by $\mathcal{M}$.

% \begin{figure}
%     \centering
%      \includegraphics[width=0.85\linewidth]{./figure/layer_mp.png}
%      \vspace{-0.3cm}
%      \caption{\red{Can delete. }The propagation paths and intra-layer processes of the general GNN model. }
%      \label{fig:massage_passing}
%      \vspace{-0.4cm}
% \end{figure}

% https://arxiv.org/pdf/1603.05027.pdf

\begin{figure*}[t]
	\centering
	%  % \hspace{-0.04\textwidth}
	\begin{subfigure}[b]{0.44\textwidth}
		\centering
		\includegraphics[width=0.99\textwidth]{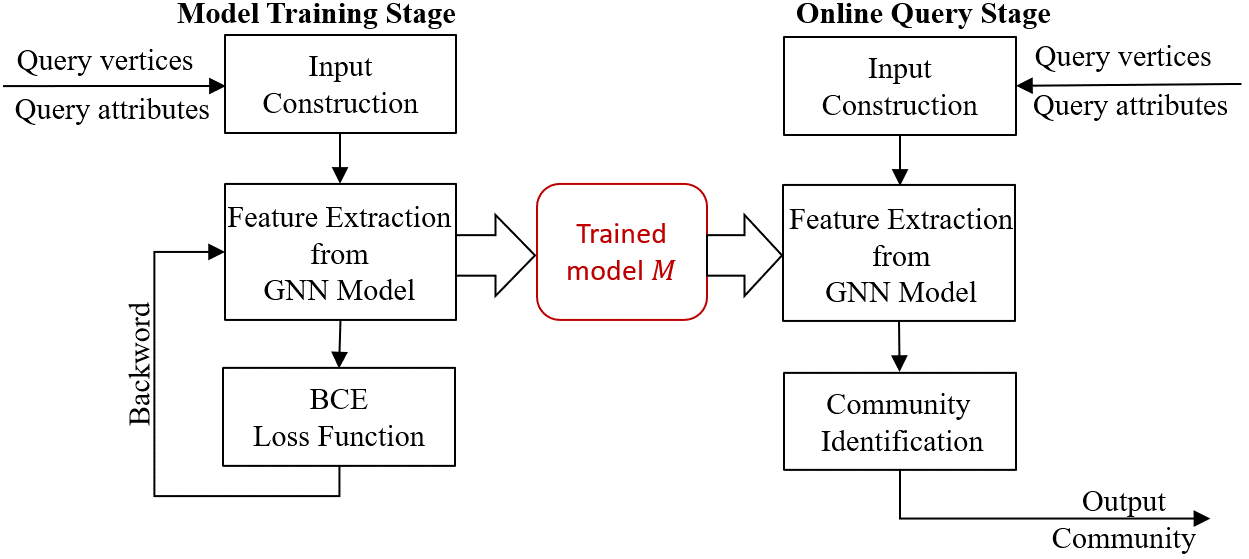}
		\vspace{-0.15CM}
		\caption{Framework of our proposed models.}
		\label{fig:framework-a}
	\end{subfigure}
	\begin{subfigure}[b]{0.44\textwidth}
		\centering
		\includegraphics[width=0.92\textwidth]{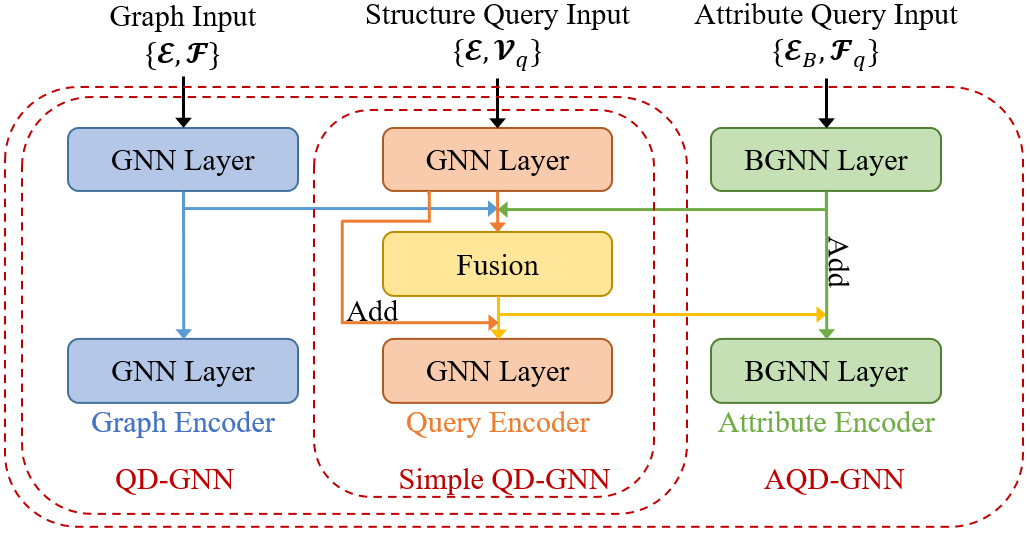}
	       \vspace{-0.15CM}
		\caption{Intermediate design of model $\mathcal{M}$ with four components.}
		\label{fig:framework-b}
	\end{subfigure}
	\vspace{-0.35CM}
	\caption{The architecture of proposed models.}
	\label{fig:framework}
	\vspace{-0.3CM}
\end{figure*}

% {\color{red} emphasize why three components }
\vspace{-0.25cm}
\subsection{A General GNN Model}
\label{Sec.generalGNN}
\vspace{-0.05cm}

We introduce a general framework of Graph Neural Network (GNN) as the cornerstone of our models.%, which also serves as a basis for most existing advanced GNN models. 

A GNN layer is known as a message passing procedure from neighborhoods. After the linear transformation of neighbors' hidden features, there are many alternative techniques within one layer, e.g., batch normalization technique~\cite{ioffe2015batch}. 
We list one of the possible intra-layer processes in the layer-wise propagation function as:
% A GNN layer is known as a message passing procedure from neighborhoods. After the linear transformation of neighbors' hidden features, there are many alternative techniques within one layer. Note that the residual (Res-Net \cite{he2016deep}) and jumping knowledge (Dense-Net \cite{huang2017densely}) are inter-layer procedures which connect multi-layer knowledge. We list one of the possible intra-layer processes in Figure~\ref{fig:massage_passing}. Correspondingly, the layer-wise propagation function can be formally defined as:
\vspace{-0.15cm}
\begin{equation}
\begin{aligned}
\boldsymbol{h}_v^{(l+1)}=\text{Dr}\left\{\phi\left(\text{BN}[\text{AGG}(\boldsymbol{h}_u^{(l)}\boldsymbol{W}^{(l+1)}+\boldsymbol{b}^{(l+1)},\mathrm{u}\in \mathcal{N}^{+}(\mathrm{v}))]\right)\right\},
\end{aligned} 
\label{eq:generalGCN}
\vspace{-0.1cm}
\end{equation}
where $\boldsymbol{h}_v^{(l+1)} \in \mathbb{R}^{d^{(l+1)}}$ is the learned new features of vertex $\mathrm{v}$ in the $(l+1)$-th layer, $\boldsymbol{h}_u^{(l)}\in \mathbb{R}^{d^{(l)}}$ is the hidden features of vertex $\mathrm{u}$ from the $l$-th layer, and the input feature $\boldsymbol{h}_v^{(0)}\in \mathbb{R}^{d}$ is the normalized form of attribute vector $\boldsymbol{f}_v$. $\boldsymbol{W}^{(l+1)}\in \mathbb{R}^{d^{(l)} \times d^{(l+1)}}$ and $\boldsymbol{b}^{(l+1)}\in \mathbb{R}^{d^{(l+1)}}$ are trainable weights. $\text{AGG}(\cdot)$ is an aggregation function such as SUM, MAX, or MIN. $\text{BN}(\cdot)$ is batch normalization \cite{ioffe2015batch} that reduces internal covariate shift. $\phi(\cdot)$ is the non-linear activation function, such as $\text{ReLU}(\cdot)$. Last, $\text{Dr}(\cdot)$ is the dropout method~\cite{srivastava2014dropout} to dilute the data and reduce the overfitting in neural networks.

% 比如 比较出名的几个 模型 他们分别的这几个函数 是啥？

For example, one of the most classical GNN models, Vanilla Graph Convolutional Network (Vanilla GCN) \cite{Kipf_GCN}, is defined as:
\vspace{-0.15cm}
\begin{equation}
\begin{aligned}
\boldsymbol{h}_v^{(l+1)}=\text{Dr}\left\{\text{ReLU}\left(\text{SUM}(\{ \frac{\boldsymbol{h}_u^{(l)}}{\sqrt{{d'_\mathrm{u}} {d'_\mathrm{v}}}} \boldsymbol{W}^{(l+1)}:\mathrm{u}\in \mathcal{N}^{+}(\mathrm{v}) \})\right)\right\},
\end{aligned} 
\label{eq:GCN_agg}	
\vspace{-0.15cm}
\end{equation}
which applies $\text{SUM}$ as the aggregation operation, and $\text{ReLU}(\cdot)$ as the activation function $\phi(\cdot)$ with the dropout method. In this GNN model, batch normalization is not adopted and Laplacian smoothing is employed where $d'_\mathrm{u}=d_\mathrm{u}+1$ and $d_\mathrm{u}$ is the degree of vertex $\mathrm{u}$.

\blue{In the following, we will focus on the way of aggregation in our proposed GNN models.  The dropout, activation function, batch normalization and the trainable bias $\boldsymbol{b}$ described above are adopted in our models, and will be omitted in our following presentation.}

\vspace{-0.2cm}
\section{The Query Framework}\label{sec.frame}

Before describing the detailed design of the proposed models, we introduce the common framework of our models for both CS and ACS problems. 
%of our models \SQD, \QD and \AQD. 
As Figure~\ref{fig:framework-a} shows, 
the proposed models consist of two main stages: \emph{the model training stage} and \emph{the online query stage}. Firstly, we train the embedding model $\mathcal{M}$ offline with the loss function in the model training stage as shown in \blue{Figure~\ref{fig:framework-a} (left)}. 
After that, in the online query stage, whenever the query comes, we apply the model from the training stage to predict the community without re-training, as \blue{Figure~\ref{fig:framework-a} (right)} presents. This framework is highly flexible. 
% By injecting different embedding models, our framework can handle different kinds of community search problems. 
In the following, we first introduce how to construct the inputs from the graph and queries in both stages. Then, we describe the two main stages respectively.

% \begin{figure*}[t]
%     \centering
%      \includegraphics[width=0.8\linewidth]{./figure/FrameA.png}
%      %\caption{The overall framework of Query-Driven GCN}
%      \vspace{-0.25CM}
%      \caption{The framework of proposed models. }
%      \label{fig:framework}
%      \vspace{-0.55CM}
% \end{figure*}

\vspace{-0.2cm}
\subsection{Input Construction}
\label{sec.frame.input}
\vspace{-0.05cm}
% Since the embedding model $\mathcal{M}$ needs the vectorized inputs. In this section, we introduce the vectorization scheme for the vertex set and attribute set.
Since the GNN model $\mathcal{M}$ needs vectorized inputs, we introduce the vectorization scheme for the vertex set and attribute set.
%As mentioned in Section~\ref{Sec.defin}, when we weed the the features into GNN models, the vertex attribute set 
%the input features of general GNN model, 
%is encoded into a matrix $\boldsymbol{F}=[\boldsymbol{f}_1, \dots, \boldsymbol{f}_n]^T \in \mathbb{R}^{n \times d}$, where $\boldsymbol{f}_v \in \mathbb{R}^d$ is the attribute vector of vertex $\mathrm{v}$. 
%In our proposed models, as Figure~\ref{fig:framework} shows, we also input the query information into the GNN models directly. Thus we need to encode both query vertices and query attributes

\vspace{-0.15cm}
\stitle{Construct query vertices. }
We encode each query vertex set $\mathcal{V}_q \subseteq \mathcal{V}$ to a one-hot vector $\boldsymbol{v}_q \in \{0,1\}^{n}$. For a query $\mathcal{V}_q$, if vertex $\mathrm{v}_i \subseteq \mathcal{V}_q$, ${\boldsymbol{v}_q}_i=1$; otherwise, ${\boldsymbol{v}_q}_i=0$. 
For example, when querying the community of vertex $\mathrm{v}_6$ in Figure~\ref{fig:example1}, the encoded vector is $\boldsymbol{v}_q=[0,0,0,0,0,1,0,0]^T$. 

\vspace{-0.15cm}
\stitle{Construct query attributes.}
Similar to query vertices, we encode each query attribute set $\mathcal{F}_q \subseteq \hat{\mathcal{F}}$ to a one-hot vector $\boldsymbol{f}_q \in \{0,1\}^{d}$, where $d=|\hat{\mathcal{F}}|$ is the number of unique attributes. %For an attribute query $\mathcal{F}_q$, if attribute $\mathrm{f}_i \subseteq \mathcal{F}_q$, the value ${\boldsymbol{f}_q}_i=1$; otherwise, ${\boldsymbol{f}_q}_i=0$. 

%For example, when attribute query is $\mathcal{F}_q=\{\text{GNN}\}$, the one-hot vector is $\boldsymbol{f}_q=[0,0,0,0,1,0,0]^T$ according to the order from A to G in Figure~\ref{fig:bipart_G_example}. 

% The encoded query vertex set, vertex attribute set and query attribute set are submitted to Query Encoder, Graph Encoder and Attribute Encoder as input features respectively.

The encoded query vertex set and query attribute set are then submitted to our proposed GNN models as input features.

\vspace{-0.25cm}
\subsection{Model Training Stage}
\vspace{-0.1cm}

In the model training stage, with a set of training queries as input, we \blue{iteratively} train the embedding model $\mathcal{M}$ offline through the Binary Cross Entropy (BCE) loss function and obtain a trained model for the online query stage. 

\blue{Given a set of training queries $\mathcal{Q}_{\text{train}}=\{q_1, q_2, ...\}$ and corresponding ground-truth communities $\mathcal{C}_{\text{train}}=\{{\mathcal{C}_{GT}}_1,{\mathcal{C}_{GT}}_2,$ $...\}$, we train a GNN model $\mathcal{M}$ to minimize the loss function to fit the training data. \blue{Given a validation query set $\mathcal{Q}_{\text{val}}$ and corresponding ground-truth communities $\mathcal{C}_{\text{val}}$, we select the parameters of model $\mathcal{M}$ and threshold $\gamma \in [0,1]$ which achieve the best performance in the validation set.}
The queries in $\mathcal{Q}_{\text{train}}$ and $\mathcal{Q}_{\text{val}}$ can be attributed $q=\{\mathcal{V}_q,\mathcal{F}_q\}$ for ACS or non-attributed $q=\{\mathcal{V}_q\}$ for CS. 

First, we construct all query inputs as one-hot vectors.  Then we repeatedly input queries into the model $\mathcal{M}$, i.e., \SQD, \QD or \AQD, which will be introduced in Section~\ref{sec.CS-GCN} and Section~\ref{sec.ACS-GCN}.  With the model $\mathcal{M}$'s output $\boldsymbol{h}_q$ for each query $q$ in an iteration, we compute BCE loss function and gradients of the model parameters. 
The gradients are propagated backward to update  $\mathcal{M}$ at the end of this iteration. 
With the updated parameters, $\mathcal{M}$ moves to the next iteration, outputs $\boldsymbol{h}_q$, calculates loss and back propagates gradients until convergence. }
The loss function of the three proposed models is the same and we describe it formally in the following.

\vspace{-0.15cm}
\stitle{Loss Function}. 
\label{Sec.pre.loss}
We formulate community search as a binary classification problem.  Assume that $\boldsymbol{h}_q \in \mathbb{R}^{n}$ is the output of $\mathcal{M}$ for query $q$ after the Sigmoid function $\sigma(x)=\frac{1}{1+e^{-x}}$, where ${{\boldsymbol{h}_q}_v}\in {[0,1]}$ represents the output for vertex $\mathrm{v}$. $\boldsymbol{y}_q \in {\{0,1\}}^{n}$ represents the ground-truth vector for query $q$. ${\boldsymbol{y}_q}_v=1$ if and only if vertex $\mathrm{v} \in {\mathcal{C}_{GT}}_q$; otherwise, ${\boldsymbol{y}_q}_v=0$. 
Then we utilize Binary Cross Entropy (BCE) function as the loss function to minimize the BCE between the model output $\boldsymbol{h}_q$ and the ground-truth label $\boldsymbol{y}_q$ for $q$.
%
% We propose a supervised learning model \QD to solve this attributed community search problem. 
The optimization loss function can be formulated as:
\vspace{-0.25cm}
\begin{equation}
    \begin{aligned}
        \min \mathcal{L}= \sum_{q \in \mathcal{Q}_{\text{train}}} \frac{1}{n} \sum_{i=1}^{n} -(\boldsymbol{y}_{q_i}\log(\boldsymbol{h}_{q_i})+(1-\boldsymbol{y}_{q_i})\log(1-\boldsymbol{h}_{q_i})). \\
        % &BCE(y_{qi},z_{qi})=\left\{
        % \begin{array}{ll}
        % -\log(z_{qi}), & {y_{qi}=1} \\
        % -\log(1-z_{qi}),  & {y_{qi}=0}
        % \end{array}
        % \right. 
    \end{aligned}
    \label{eq:loss}	
\vspace{-0.2cm}
\end{equation}
% In \eqref{eq:loss}, $\mathcal{Q}_{\text{train}} = \{(\mathcal{V}_q, \mathcal{F}_q, y_q)\}_{q=1}^{k}$ is the training set of queries. The triplet $(\mathcal{V}_q, \mathcal{F}_q, y_q)$ represents a query $\langle \mathcal{V}_q,\ \mathcal{F}_q \rangle$ and its corresponding ground-truth community denoted by the indicator vector $y_q\in {\{0,1\}}^{n \times 1}$.  If $v_i$ is in the ground truth community, $y_{qi}=1$; otherwise, $y_{qi}=0$. $z_q\in \mathbb{R}^{n \times 1}$ is the output prediction vector by \QD for query $q$.

\vspace{-0.3cm}
\subsection{Online Query Stage} 
\label{sec.frame.onlineQuery}
\vspace{-0.1cm}

\blue{In the online query stage, we utilize the well-trained model $\mathcal{M}$ and threshold $\gamma$ from the model training stage to process the online query $q$ and produce the community $\mathcal{C}_q$ without re-training.  We first construct query inputs as one-hot vectors. 
Then the constructed vectors are fed into model $\mathcal{M}$, which only runs once and outputs the vector $\boldsymbol{h}_q$.  To ensure the connectivity between query vertices and community members, we employ a constrained Breadth-First Search (BFS) starting from the query vertices in Algorithm~\ref{algo:BFS}. When visiting vertex $\mathrm{u}$, if $\boldsymbol{h}_{q_u} \geq \gamma$ (line 4), we add vertex $\mathrm{u}$ to the output community $\mathcal{C}_q$ (line 6).
}

\blue{
Please note that the connectivity of the output community also depends on the user-specified query vertices. If the induced subgraph of the query vertices is connected, then our models are guaranteed to find a connected community. If the induced subgraph of the query vertices is not connected, our models may still find a connected community through some bridging vertices. But there is possibility that the discovered community is not connected as one component, especially when the query vertices are distant or disconnected in the graph. In this case, our models can still find some connected
components, each of which contains part of the query vertices, as the answer community.
}

\begin{algorithm}[t]
\footnotesize
\caption{\blue{Constrained BFS for Community Identification}} \label{algo:BFS}
\blue{
\begin{flushleft}
\footnotesize
\textbf{Input:} Graph: $G=(\mathcal{V}, \mathcal{E})$, a query vertex set: $\mathcal{V}_q$, \\
\hspace{1.1cm} a model output vector: $\boldsymbol{h}_q$, a threshold: $\gamma$.\\
\textbf{Output:} a vertex set of community : ${\mathcal{C}_q}$.\\
\end{flushleft}
%\vspace{0.1cm}
\begin{algorithmic}[1]
\footnotesize
% \STATE  Build \QD $\mathcal{M}$ with two branches SE ($\boldsymbol{h}_Q$) and GE ($\boldsymbol{h}_G$), and a fusion operator ($\boldsymbol{h}_{FF}$).

\STATE  Initialize set ${\mathcal{Q}}=\mathcal{V}_q$, ${\mathcal{C}_q}=\mathcal{V}_q$

\STATE  \textbf{while} $\mathcal{Q}$ is not empty \textbf{do}

\STATE  \hspace{0.3cm} select a vertex $\mathrm{v}$ from $\mathcal{Q}$

\STATE  \hspace{0.3cm} \textbf{for} $\mathrm{u} \in \mathcal{N}(\mathrm{v})$ and ${\boldsymbol{h}_q}_u \geq \gamma$ \textbf{do}

\STATE  \hspace{0.3cm} \hspace{0.3cm} $\mathcal{Q} \leftarrow {\mathcal{Q}} \cup \{\mathrm{u}\}$

\STATE  \hspace{0.3cm} \hspace{0.3cm} ${\mathcal{C}_q} \leftarrow {\mathcal{C}_q} \cup \{\mathrm{u}\}$

\STATE  \textbf{return} ${\mathcal{C}_q}$;
\end{algorithmic}
}
\end{algorithm}
% \vspace{-0.15cm}

% \blue{Del: It's worth to note that our framework only runs once to predict the community for online query, while the state-of-the-art learning-based model, ICS-GNN needs to re-train the model for each query. It is hard handle the online-query for large graphs with strict time limits.}

%\vspace{-0.1cm}
%\stitle{Community Identification}. 
%Suppose \SQD, \QD or \AQD is a three layer model. Since community search problem is formed as multiple binary classification task in Section~\ref{Sec.defin}, the last layer needs to translate hidden feature of each vertex into a value with weight matrix $\boldsymbol{W}^{(3)}\in \mathbb{R}^{d^{(2)} \times 1}$ for further classification. The output of models is ${\boldsymbol{h}}^{(3)} \in {[0,1]}^{n}$ after the Sigmoid function, where ${\boldsymbol{h}_v}\in {[0,1]}$ represents the output for vertex $\mathrm{v}$. We utilize a learnable threshold $\gamma$ and query connectivity in graph to convert the ${\boldsymbol{h}}^{(3)}$ into community $\mathcal{C}$. 
%If output ${\boldsymbol{h}_v}^{(3)} > \gamma$ and vertex $\mathrm{v}$ is connected with any query vertices in graph, vertex $\mathrm{v}$ belong to the output community $\mathcal{C}$; otherwise, $\mathrm{v} \notin \mathcal{C}$.

% \section{Preliminaries}\label{sec.pre}
% \input{pre}

% \vspace{-0.2cm}
\section{QD-GNN Model for CS}\label{sec.CS-GCN}
\vspace{-0.1cm}
%In this section, we develop GNN-based models to tackle the community search problem. 
In this section, we introduce the construction of the embedding model $\mathcal{M}$ in the proposed framework for community search. We first propose a Simple task-oriented Query Driven-Graph Neural Network (Simple \QD) and then design useful functional encoders to improve it as \QD model. As Figure~\ref{fig:framework-a} shows, with query vectors as input, the \SQD or \QD model $\mathcal{M}$ outputs $\boldsymbol{h}_q$ into the BCE loss function during the training process. In the online query stage, the model output $\boldsymbol{h}_q$ is translated into community members as described in Section~\ref{sec.frame}.

\vspace{-0.1cm}
\subsection{Simple QD-GNN}
\label{Sec.SimpleCS}
\vspace{-0.1cm}
% input construction
% Community Identification

%In order to learn the queries without any gap, 
%In order to learn different underlying community information for different queries, we propose a \SQD model for community search based on the general GNN as introduced in Section~\ref{Sec.generalGNN}. 

\blue{The \SQD model is designed based on the general GNN introduced in Section~\ref{Sec.generalGNN} and uses query vector $\boldsymbol{v}_q$ as the input features of the model. This model input enables query-centered structural propagation, i.e., propagating from the query vertices to its neighborhood, to better capture the local query structure information.
}

%Reconsider the function in Eq.~\eqref{eq:generalGCN}, GNN propagates the attribute information of vertices through graph structures. 
%\blue{However, previous algorithms for community search problem  \cite{sozio2010community, cui2013online, cui2014local, huang2014querying, huang2015approximate, chang2015index, hu2016querying, akbas2017truss, yuan2017index} do not utilize vertex attributes, which only focus on local query structures in non-attributed graphs}
% However, online community search problem focuses on local query structures in non-attributed graphs, leading to that the corresponding previous algorithms do not utilize vertex attributes neither. 
%It thus becomes natural to build a \QD model and replaces attributes by query information as input features in Eq.~\eqref{eq:generalGCN} to capture local query structure information. 
%In Input Construction, we encode each query vertex set $\mathcal{V}_q \subseteq \mathcal{V}$ to an one-hot vector $\boldsymbol{v}_q \in \{0,1\}^{n}$ as introduced in Section~\ref{sec.frame.input}. 
% For a query $\mathcal{V}_q$ in community search problem, if vertex $\mathrm{q}_i \subseteq \mathcal{V}_q$, the value ${\boldsymbol{v}_q}_i=1$; otherwise, ${\boldsymbol{v}_q}_i=0$. 
%We take query vector $\boldsymbol{v}_q$ as input features of GNN model, i.e., $\boldsymbol{h}_u^{(0)}={\boldsymbol{v}_q}_u$ and the query knowledge propagates through edges for the non-attributed community search problem. 

We name this query driven propagation as \emph{Graph Encoder}. In order to fully make use of vertex \blue{features in each layer}, Query Encoder is designed to equip with a self feature modeling \cite{self_feat}. The inter-layer propagation function for vertex $\mathrm{v}$ is formally defined as:
% \begin{align}
% H_S^{(l+1)} = \hat{A}I_S^{(l)}W_{S}^{(l)} + {H_{S}}^{(l)}W_{S_\text{self}}^{(l)},
% \label{eq:SE-encoder}
% \end{align}
\vspace{-0.15cm}
\begin{equation}
\begin{aligned}
\boldsymbol{h}_{Q_v}^{(l+1)}=\boldsymbol{h}_{Q_v}^{(l)} \boldsymbol{W}_{Q_\text{self}}^{(l+1)} + \text{SUM}(\{\boldsymbol{h}_{Q_u}^{(l)} \boldsymbol{W}_Q^{(l+1)} : \mathrm{u}\in \mathcal{N}^+(\mathrm{v})\}),
\end{aligned} 
\label{eq:SE-encoder}	
\vspace{-0.05cm}
\end{equation}
\blue{where the first component emphasizes the self features (hidden features of the vertex $\mathrm{v}$) with learnable weight parameter matrices $W_{Q_\text{self}}^{(l+1)}\in \mathbb{R}^{d^{(l)} \times d^{(l+1)}}$. The second component is similar to Eq.~\eqref{eq:generalGCN} with a subscript $Q$, and chooses SUM as the aggregation function as Vanilla GCN \cite{Kipf_GCN} does. 
Similarly, $\boldsymbol{W}_Q^{(l+1)}\in \mathbb{R}^{d^{(l)} \times d^{(l+1)}}$ is the trainable weight matrix , $\boldsymbol{h}_{Q_v}^{(l+1)} \in \mathbb{R}^{d^{(l+1)}}$ is the learned new features of vertex $\mathrm{v}$ in the $(l+1)$-th layer of Query Encoder. 
Different from Eq.~\eqref{eq:generalGCN}, the input feature of the first layer $\boldsymbol{h}_{Q_v}^{(0)}$ is the one-hot query vector $\boldsymbol{v}_{q_v}$.
% notations are similar to Eq.~\eqref{eq:generalGCN} with a subscript $Q$, and $W_{Q_\text{self}}^{(l+1)}\in \mathbb{R}^{d^{(l)} \times d^{(l+1)}}$ are the weight parameter matrices. The difference lies on the input feature of the first layer $\boldsymbol{h}_{Q_v}^{(0)}$, which is the one-hot query vector $\boldsymbol{v}_{q_v}$ from Input Construction. 
}

\begin{figure}[t]
% \small
    \centering
    \vspace{-0.2cm}
     \includegraphics[width=0.75\linewidth]{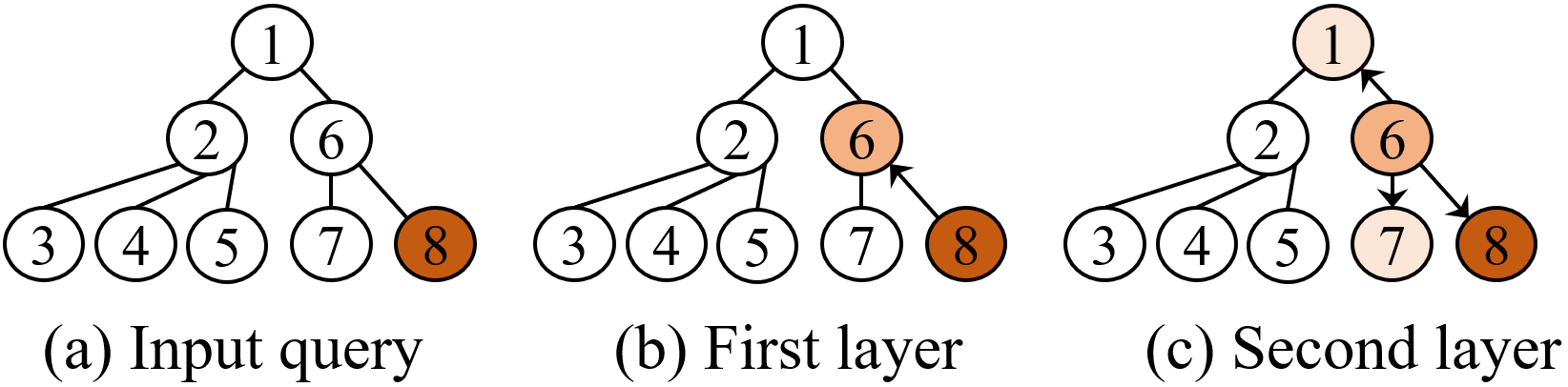}
     \vspace{-0.5cm}
     \caption{Query propagation paths in Query Encoder. }
     \label{fig:query_propagation}
     \vspace{-0.4cm}
\end{figure}

\vspace{-0.15cm}
\begin{example}
We follow the example in Figure~\ref{fig:example1} and show the propagation paths in Figure~ \ref{fig:query_propagation}. For query $\mathcal{V}_q=\{\mathrm{v}_8\}$ highlighted in Figure~\ref{fig:query_propagation}a, the query vector ${\boldsymbol{v}_q}$ is $[0,0,0,0,0,0,$ $0,1]^T$. According to Eq.~\eqref{eq:SE-encoder}, in the first layer, the query information propagates to the neighbor of $\mathrm{v}_8$, i.e., $\mathrm{v}_6$ as depicted in Figure~\ref{fig:query_propagation}b. Then, the 2-hop neighbors of the query vertex, $\mathrm{v}_1$ and $\mathrm{v}_7$, acquire the knowledge from $\mathrm{v}_6$ in the second layer as depicted in Figure~\ref{fig:query_propagation}c.
\end{example}
\vspace{-0.15cm}

% The output of \SQD in the last layer $\boldsymbol{h}_Q$ 
% is submitted to BCE loss function in the model training stage, and transferred to the predicted community in the online query stage, as described in Section~\ref{sec.frame}.
% \stitle{Community Identification}. Suppose \SQD is a three layer model. Since community search problem is formed as multiple binary classification task in Section~\ref{Sec.defin}, the last layer need to translate hidden feature of each vertex into a value with weight matrix $\boldsymbol{W}^{(3)}\in \mathbb{R}^{d^{(2)} \times 1}$ for further classification. The output of \SQD is ${\boldsymbol{h}}^{(3)} \in {[0,1]}^{n}$ after the Sigmoid function, where ${\boldsymbol{h}_v}\in {[0,1]}$ represents the output for vertex $\mathrm{v}$. We utilize a learnable threshold $\gamma$ to convert the ${\boldsymbol{h}}^{(3)}$ into community $\mathcal{C}$. 
% If output ${\boldsymbol{h}_v}^{(3)} > \gamma$, vertex $\mathrm{v}$ is belong to the output community $\mathcal{C}$. Otherwise, $\mathrm{v} \notin \mathcal{C}$.

\vspace{-0.2cm}
\subsection{QD-GNN}
\label{Sec.CS.CS-GNN}
\vspace{-0.05cm}
% \begin{figure}[t]
%     \centering
%      \includegraphics[width=0.9\linewidth]{./figure/CS-GNN2.png}
%      \vspace{-0.13in}
%      \caption{The framework of \QD with two branches: Graph Encoder and Query Encoder. These two branches are combined with a fusion operation and final output community vector $z_q$ is the fusion result in the last layer.}
%      \label{fig:CS-GNN}
%      \vspace{-15pt}
% \end{figure}

%GNN models achieve great success
%GNN models have been widely adopted to learn vertex attributes and network structure simultaneously in many graph analytic tasks. Most recently, ICS-GNN~\cite{gao2021ics_ICS-GNN} develops a GNN-based learning approach to  address the interactive community search problem, where the vertex attributes are utilized to train the learning model but only query vertices with no query attributes are issued. 

\blue{
Recent studies \cite{GraphZoom, wang2020amgcn, gao2021ics_ICS-GNN} have found that attributes on graph vertices can be leveraged for structural learning problems, for example, link prediction \cite{GraphZoom} and community search \cite{gao2021ics_ICS-GNN}.  Inspired by their findings, we design an improved \QD model based on \SQD and Vanilla GCN \cite{Kipf_GCN}.  Similar to ICS-GNN~\cite{gao2021ics_ICS-GNN}, \QD combines the network structure and vertex attributes to solve the community search problem.}

% Other than encoding the query information only, ICS-GNN~\cite{gao2021ics_ICS-GNN} suggests that considering the whole graph can further boost the community search performance and improve the model generalization ability for the unseen queries. 
% Following this suggestion, we develop an improved \QD model by learning both network structure and vertex attributes, leveraging the \SQD in Section~\ref{Sec.SimpleCS} and Vanilla GCN \cite{Kipf_GCN} in Section~\ref{Sec.generalGNN}. 

\vspace{-0.15cm}
\subsubsection{Overview}
% The whole algorithm of \QD-based community search is presented in Algorithm~\ref{algo:TruDec}. We train the \QD model offline only once and answer any possible online query vertices $\mathcal{V}_q \subseteq \mathcal{V}$ for community search in a stream setting (lines 1-5). The procedure of \QD model construction is also depicted (lines 6-23). 
% Specifically, 
%As Figure~\ref{fig:framework-a} shows, with inputs from Input Construction, \QD model outputs $\boldsymbol{h}_q$ into the BCE loss function during the training process. While, in the online query, Community Identification translates model output $\boldsymbol{h}_q$ into community members with a threshold $\gamma$ as described in Section~\ref{sec.frame}.

Figure~\ref{fig:framework-b} presents the architecture overview of \QD model, which consists of two convolution branches (\emph{Graph Encoder} and \emph{Query Encoder}) and a \emph{Feature Fusion} operator.  
\emph{Graph Encoder} provides the query-independent information with both graph structure and vertex attributes as input, i.e., the edge set $\mathcal{E}$ and vertex attribute set $\mathcal{F}$.  
\emph{Query Encoder} (the same as that in \SQD) provides the interface for query vertices and learns the query-specific local topology features. It takes the input of graph structure and query vertices, i.e., the edge set $\mathcal{E}$ and query vertices $\mathcal{V}_q$. 
The \emph{Feature Fusion} operator combines the above encoder embedding results and obtains the final query-specific output vectors. This fusion makes use of both global graph knowledge and local query information which can achieve a good balance, and finally obtains the model output $\boldsymbol{h}_q$ for each query $q$.

\vspace{-0.15cm}
\subsubsection{Graph Encoder} Graph Encoder focuses on global graph structure and vertex attributes, both of which are independent of queries. We apply the layer-wise forward propagation of the general GNN to construct Graph Encoder, which has been introduced in Section~\ref{Sec.generalGNN}. 
% In order to fully make use of vertex attributes, Graph Encoder is designed to equip with a self feature modeling \cite{self_feat}. 
Similar to \SQD, the forward layer of Graph Encoder is defined with a self feature modeling \cite{self_feat} as:
% \begin{equation}
% \begin{aligned}
% H_G^{(l+1)} = \hat{A}H_G^{(l)}W_{G}^{(l)} + H_G^{(l)}W_{G_\text{self}}^{(l)},
% \end{aligned}
% \label{GE-encoder}	
% \end{equation}
\vspace{-0.1cm}
\begin{equation}
\begin{aligned}
\boldsymbol{h}_{G_v}^{(l+1)}=\boldsymbol{h}_{G_v}^{(l)} \boldsymbol{W}_{G_\text{self}}^{(l+1)}+\text{SUM}(\{ \boldsymbol{h}_{G_u}^{(l)} \boldsymbol{W}_G^{(l+1)}:{\mathrm{u}\in \mathcal{N}^+(\mathrm{v})}\}),%\mathrm{u}\in \mathcal{N}^+(\mathrm{v}),
\end{aligned} 
\label{eq:GE-encoder}	
\vspace{-0.1cm}
\end{equation}
where the notations are the same as Eq.~\eqref{eq:generalGCN} with a subscript $G$, and $W_{G_\text{self}}^{(l+1)}\in \mathbb{R}^{d^{(l)} \times d^{(l+1)}}$ are the weight parameter matrices. The input feature of vertex $\mathrm{v}$ in the first layer $\boldsymbol{h}_{G_v}^{(0)}\in \mathbb{R}^{d}$ is the normalized attribute vector $\boldsymbol{f}_v$ encoded in Section~\ref{Sec.defin}. Graph Encoder propagates the attribute information through graph structure and learns query-independent knowledge.

\vspace{-0.15cm}
\begin{example}
We follow the example in Figure \ref{fig:example1} to illustrate how Graph Encoder works. For vertex $\mathrm{v}_8$, in the first layer, its attributes (``DL'' and ``CV'') are propagated to its neighbor, vertex $\mathrm{v}_6$, with a learnable weight. At the same time, the attribute of vertex $\mathrm{v}_6$ (``ML'') is also propagated to vertex $\mathrm{v}_8$. In the next layer, those attributes are propagated to their neighbors respectively as well. By this propagation, the attributes of vertices $\mathrm{v}_6$, $\mathrm{v}_7$ and $\mathrm{v}_8$ become more similar.  This information is used by the Feature Fusion operator to identify the community members more accurately.  
% In contrast, \SQD does not use the global graph structure and vertex attributes for model learning.
\end{example}

\vspace{-0.25cm}
\subsubsection{Query Encoder}
Query Encoder is the same as that of \SQD and provides an interface for query vertices and obtains the local structure knowledge.  Inputs of the Query Encoder are based on the graph topology (graph edges $\mathcal{E}$) and structural query (query vertices $\mathcal{V}_q$).
% Define $I_S^{(l)}$ as the input feature of Query Encoder in the $(l+1)$-th layer and $H_S^{(l+1)}$ as the output feature. 
The inter-layer propagation function of Query Encoder is the same as that in Eq.~\eqref{eq:SE-encoder}.

\begin{algorithm}[t]
\footnotesize
\caption{The $k$-Layer \QD Propagation} \label{algo:QD}
\begin{flushleft}
\footnotesize
\textbf{Input:} Graph: $G=(\mathcal{V}, \mathcal{E}, \mathcal{F})$, \\
\hspace{0.75cm} a set of queries: $\mathcal{Q}=\{{\mathcal{V}_q}_1,{\mathcal{V}_q}_2,\dots\}$, \\
\hspace{0.75cm} \QD model: $\mathcal{M}=\{\boldsymbol{h}_Q, \boldsymbol{h}_G, \boldsymbol{h}_{FF}\}$.\\
\textbf{Output:} a set of output vectors: $\mathcal{H}=\{{\boldsymbol{h}_q}_1,{\boldsymbol{h}_q}_2,\dots\}$.\\
\end{flushleft}
%\vspace{0.1cm}
\begin{algorithmic}[1]
\footnotesize
% \STATE  Build \QD $\mathcal{M}$ with two branches SE ($\boldsymbol{h}_Q$) and GE ($\boldsymbol{h}_G$), and a fusion operator ($\boldsymbol{h}_{FF}$).

\STATE  Construct attribute matrix ${\boldsymbol{F}}$ for ${\mathcal{F}}$

\STATE  $\mathcal{H} \leftarrow \varnothing$

\STATE  \textbf{for} each ${\mathcal{V}_q}\in \mathcal{Q}$ \textbf{do}

\STATE  \hspace{0.3cm} Construct one-hot vector ${\boldsymbol{v}_q}$ for ${\mathcal{V}_q}$, initialize $\boldsymbol{h}_Q^{(0)}$ with ${\boldsymbol{v}_q}$

\STATE  \hspace{0.3cm} Initialize $\boldsymbol{h}_G^{(0)}$ with $\boldsymbol{F}$

\STATE  \hspace{0.3cm} $\boldsymbol{h}_Q^{(1)} \leftarrow \text{Propg}(\boldsymbol{h}_{Q}^{(0)}, \mathcal{E})$ in Eq.~\eqref{eq:SE-encoder}%Propagate ${\boldsymbol{h}_{S}}^{(0)}$ through edges in $\mathcal{E}$ and obtain ${\boldsymbol{h}_Q}^{(1)}$ (Eq.~\eqref{eq:SE-encoder2})

\STATE  \hspace{0.3cm} $\boldsymbol{h}_G^{(1)} \leftarrow \text{Propg}(\boldsymbol{h}_{G}^{(0)}, \mathcal{E})$ in Eq.~\eqref{eq:GE-encoder}%Propagate ${\boldsymbol{h}_{G}}^{(0)}$ through edges in $\mathcal{E}$ and obtain ${\boldsymbol{h}_G}^{(1)}$ (Eq.~\eqref{eq:GE-encoder})

\STATE  \hspace{0.3cm} $\boldsymbol{h}_{FF}^{(1)} \leftarrow \AGG(\boldsymbol{h}_{G}^{(1)},\boldsymbol{h}_{Q}^{(1)})$ in Eq.~\eqref{eq:FuseCS-GCN}

\STATE  \hspace{0.3cm} $l \leftarrow 1$

\STATE  \hspace{0.3cm} \textbf{while}($l<k$) \textbf{do}

\STATE  \hspace{0.3cm} \hspace{0.3cm} $\boldsymbol{h}_Q^{(l+1)} \leftarrow \text{Propg}(\boldsymbol{h}_{FF}^{(l)}, \mathcal{E})$ in Eq.~\eqref{eq:SE-encoder2}%Propagate {${\boldsymbol{h}_{FF}}^{(l)}$} through edges in $\mathcal{E}$ and obtain ${\boldsymbol{h}_Q}^{(l+1)}$ (Eq.~\eqref{eq:SE-encoder2})

\STATE  \hspace{0.3cm}  \hspace{0.3cm} $\boldsymbol{h}_G^{(l+1)} \leftarrow \text{Propg}(\boldsymbol{h}_{G}^{(l)}, \mathcal{E})$ in Eq.~\eqref{eq:GE-encoder}%Propagate ${\boldsymbol{h}_{G}}^{(l)}$ through edges in $\mathcal{E}$ and obtain ${\boldsymbol{h}_G}^{(l+1)}$ (Eq.~\eqref{eq:GE-encoder})

\STATE  \hspace{0.3cm} \hspace{0.3cm} $\boldsymbol{h}_{FF}^{(l+1)} \leftarrow \AGG(\boldsymbol{h}_{G}^{(l+1)},\boldsymbol{h}_{Q}^{(l+1)})$ in Eq.~\eqref{eq:FuseCS-GCN}

\STATE  \hspace{0.3cm} \hspace{0.3cm} $l \leftarrow l+1$

% \STATE  \hspace{0.3cm} Translate ${\boldsymbol{h}_{FF}}^{(3)}$ into vertex set $\mathcal{C}_q$ with threshold $\gamma$

\STATE  \hspace{0.3cm} $\mathcal{H} \leftarrow \mathcal{H} \cup \boldsymbol{h}_{FF}^{(k)}$   % union or add? append ?

\STATE  \textbf{return} $\mathcal{H}$;

\end{algorithmic}
\end{algorithm}

\vspace{-0.25cm}
\subsubsection{Feature Fusion}
The Feature Fusion operator combines output features learned by the above two encoders, and balances the global and local information to get the final output of \QD. The inputs of Feature Fusion are based on the output of the two encoders, i.e., $\boldsymbol{h}_G$ and $\boldsymbol{h}_Q$. It fuses them and transmits the fusion result to Query Encoder as shown in Figure~\ref{fig:framework-b}.

Based on the output of the two encoders, the forward layer of Feature Fusion is formulated as:
\vspace{-0.1cm}
\begin{equation}
\begin{aligned}
\boldsymbol{h}_{{FF}_v}^{(l+1)}=\AGG(\boldsymbol{h}_{G_v}^{(l+1)}, \boldsymbol{h}_{Q_v}^{(l+1)}),
\end{aligned}
\label{eq:FuseCS-GCN}
\vspace{-0.2cm}
\end{equation}
where $\boldsymbol{h}_{{FF}_v}^{(l+1)}$ is the output of Feature Fusion for vertex $\mathrm{v}$ and also the final output of the entire \QD model in the $(l+1)$-th layer, $\AGG(\cdot)$ is the aggregation function (e.g., Concatenation, SUM, etc.), and $\boldsymbol{h}_{G_v}^{(l+1)}$, $\boldsymbol{h}_{Q_v}^{(l+1)}$ are the outputs of each encoder for vertex $\mathrm{v}$ in the $(l+1)$-th layer respectively.

For Graph Encoder, we do not use the fusion result and just use the output of Graph Encoder itself in the $l$-th layer $\boldsymbol{h}_G^{(l)}$ as the input of the $(l+1)$-th layer. \blue{Thus, we keep Graph Encoder independent of query information in the intermediate layer. This query-independent features provide stable ``prior'' knowledge about the graph and supply additional information for community search problem, which makes \QD a stronger model.
}

For Query Encoder, we replace the feature propagation between neighbors as the fusion features in the intermediate layers. This fusion operation transmits the vertex attributes and global structure features into Query Encoder and delivers these features around query vertices.
We define $\hat{\boldsymbol{h}_Q}$ as the input feature of each layer which can be formally written as:
\vspace{-0.2cm}
\begin{equation}
\begin{aligned}
 \hat{\boldsymbol{h}}_{Q_i}^{(l)}=\left\{
 \begin{array}{ll}
 {\boldsymbol{v}_q}_i, & \text{if } l = 0;\\
 \boldsymbol{h}_{{FF}_i}^{(l)}, & \text{otherwise}.\\
 \end{array}
 \right.
 \label{eq:hsl}
\end{aligned}
\vspace{-0.15cm}
\end{equation}
The propagation function of Query Encoder can be rewritten as:
\vspace{-0.15cm}
\begin{equation}
\begin{aligned}
\boldsymbol{h}_{Q_v}^{(l+1)}=\boldsymbol{h}_{Q_v}^{(l)} \boldsymbol{W}_{Q_\text{self}}^{(l+1)}+\text{SUM}(\{ \hat{\boldsymbol{h}}_{Q_u}^{(l)} \boldsymbol{W}_Q^{(l+1)}:\mathrm{u}\in \mathcal{N}^+(\mathrm{v})\}).
\end{aligned} 
\label{eq:SE-encoder2}	
\vspace{-0.15cm}
\end{equation}

\vspace{-0.1cm}
\subsubsection{Algorithm}
The \QD model for the community search problem is presented in Algorithm~\ref{algo:QD}. For easy description, we simplify the propagation function in each encoder as $\text{Propg}(\boldsymbol{h},\mathcal{E})$, which means propagating feature $\boldsymbol{h}$ through edges in $\mathcal{E}$. 
At the beginning, we construct the feature matrix for the graph (line 1) and set the output as empty (line 2).
For each query, we also construct the query vector and initialize Query Encoder and Graph Encoder (line 4-5). 
In the first layer (line 6-9), Query Encoder and Graph Encoder propagate their input features through the graph edges (line 6-7), and Feature Fusion fuses the output of them (line 8). 
In the intermediate layers (line 10-14), Query Encoder utilizes the fused feature from Feature Fusion (line 11), while Graph Encoder takes its own output $\boldsymbol{h}_G$ as the input feature to remain independent of the query (line 12). 
The final output of \QD is the fused feature $\boldsymbol{h}_{FF}$ and we add it into the output set $\mathcal{H}$ (line 15). 

%As the framework introduced in Section~\ref{sec.frame}, in the model training stage, the output set $\mathcal{H}$ is submitted to BCE loss function in Eq.~\eqref{eq:loss} and optimizes the model through backward steps. In the online query stage, the output set is transferred to communities for each online query through Community Identification in Section~\ref{sec.frame.onlineQuery}.

% \vspace{-0.2cm}
\section{AQD-GNN Model for ACS}\label{sec.ACS-GCN}
% \section{ACS-GCN}
In this section, we extend \QD by incorporating the query attributes and propose the GNN model for attributed community search, named Attributed Query Driven-Graph Neural Network (\AQD). We first identify the challenges of attributed community search when using GNN models. Then, we describe the components of \AQD one by one in detail. 

% \begin{figure*}
%     \centering
%      \includegraphics[width=0.8\linewidth]{./figure/layer1.pdf}
%      %\caption{The overall framework of Query-Driven GCN}
%      %\caption{The framework of QD-GCN}
%      \caption{An illustration of three-layer \QD framework. Given the graph adjacent matrix $A$, the attribute matrix $F$, the query $<\mathcal{V}_q, \mathcal{F}_q>$, \QD employs three GCN-based encoders and one fusion components to model different aspects of the attributed community search task. }
%      \label{fig:layer}
%      \vspace{-3pt}
% \end{figure*}

\vspace{-0.2cm}
\subsection{Challenges}
\label{sec.ACS.chanl}
\vspace{-0.1cm}
%Given a graph $G=(A, F)$ and a query <$\mathcal{V}_q,\ \mathcal{F}_q$> as an input, our \QD model aims to identify structure cohesive and attribute homogeneous community $\mathcal{C}_q$ as an output. In terms of learning-based model, there are three tasks for attribute community search problem.

\blue{Different from community search~\cite{gao2021ics_ICS-GNN, sozio2010community, cui2014local, huang2014querying, akbas2017truss}, the attributed community search task~\cite{fang2016effective,huang2017attribute}  needs to integrate the query attributes into models. 
% which is difficult to integrate this attribute learning part into existing models. 
However, the meaning of query attributes $F_q$ and the dimension of query attributes vector $f_q$ are different from those of query vertices. It is not feasible to input query attribute information as we handle query vertices in Section~\ref{sec.CS-GCN}. }

The ICS-GNN model~\cite{gao2021ics_ICS-GNN} utilizes the query vertex information as the labels of vertices, and aligns the output embedding and the labels through a loss function. Since previous studies always focus on tasks at the level of vertices and edges, such as node classification and link prediction, but not at the attribute level for attributed queries, the design of their loss functions also centers on the vertices. The BCE loss function in Eq.~\eqref{eq:loss} is an example, which focuses on the class of each vertex. Therefore, ICS-GNN cannot incorporate the query attributes in the loss function directly and thus is not able to extend to the ACS problem.

The similar phenomenon can be observed from the \QD model, which considers query vertices as the input features and propagates the query information via edges to find local structures surrounding the query vertices.  But the query attributes cannot be easily incorporated as model input due to the different dimensionality. Even if we have a mechanism to take query attributes as input features, this attribute information can only propagate to adjacent vertices via graph topology by \QD, but cannot reach vertices having similar attributes to the query attributes, as ACS aims to do.% is to find community members sharing similar attributes to the given query attributes. 

%rather than those similar vertices in attribute. For query attributes, ACS aims to find community members shared with similar attributes to the given query attributes.

The above discussions reveal that \emph{incorporating query attributes into the learning model} and \emph{identifying the vertices with similar attributes automatically} are two key issues to be addressed in applying GNN models into the ACS problem. In \AQD, we design a bipartite graph to represent the relations between vertices and attributes. Leveraging this bipartite graph, \AQD can accept an input of query attributes and translate this query attribute knowledge into vertex knowledge. Finally, \AQD can find the vertices which have similar attributes with the query attributes.

% \QD has four input variable for attributed community search, including the adjacent matrix $A \in {\mathbb{R}}^{n*n}$, vertex feature matrix $F \in {\mathbb{R}}^{n*d}$, query vertex set $\mathcal{V}_q \subseteq \mathcal{V}$, and query attribute set $\mathcal{F}_q \subseteq \mathcal{F}$.

\vspace{-0.25cm}
\subsection{Overview}
\vspace{-0.1cm}

\AQD takes an attributed graph $G=(\mathcal{V}, \mathcal{E}, \mathcal{F})$ and a group of attributed queries 
% $q=\langle \mathcal{V}_q,\ \mathcal{F}_q \rangle$ 
$q=\langle \boldsymbol{v}_q,\ \boldsymbol{f}_q \rangle$ 
vectorized as inputs, and predicts the community vector $\boldsymbol{h}_q$ as outputs for each query $q$. 
Figure \ref{fig:framework-b} illustrates the inter-layer design of \AQD, which consists of a Feature Fusion operator and three GNN components: Graph Encoder, Query Encoder and Attribute Encoder. Note that Graph Encoder and Query Encoder are the same as those of \QD in Section~\ref{Sec.CS.CS-GNN}. Attribute Encoder is a new component specifically designed for ACS. Accordingly, Feature Fusion needs to be revised due to the new Attribute Encoder. In the following, we describe Attribute Encoder and the revised Feature Fusion operator. 

\vspace{-0.1cm}
\stitle{Attribute Encoder}. Attribute Encoder serves as the interface of query attributes and provides attribute information related to queries. It views each attribute as an individual vertex and models vertex attributes as a bipartite graph between vertex set $\mathcal{V}$ and attribute set $\hat{\mathcal{F}}$. With this bipartite graph, query attributes can be inputted into the attribute side directly and propagated between the vertex side and attribute side. Through this propagation, Attribute Encoder learns query-specific attribute node embeddings and identifies the vertices with attributes similar to queries.

% It takes the input of graph attributes and query attributes, i.e., vertex feature matrix $F$ and query attributes $\mathcal{F}_q$. Besides attribute similarity, we also consider the node-attribute relationships through a bipartite graph to obtain the query-specific attribute embedding $E_A$.
%Similarly, Attribute Encoder itself invokes no structure information and encodes attribute information to produce the attribute embedding $E_A$.
\vspace{-0.1cm}
\stitle{Feature Fusion}. The Feature Fusion component combines all the above embeddings and obtains the final query-specific output of the \ACS problem. It takes the outputs of the three encoders as inputs, mixes global graph features and local query features, fuses structure and attribute information, and balances them to get an accurate community. Note that the final output of the entire model is the fused result in the last layer.

In the following sections, we will illustrate the detailed working mechanism of Attribute Encoder and Feature Fusion.% in more detail. %The detailed descriptions of Graph Encoder and Query Encoder are presented in Section~\ref{sec.CS-GCN}.

\begin{figure}[t]
 \centering
 \includegraphics[width=0.8\linewidth]{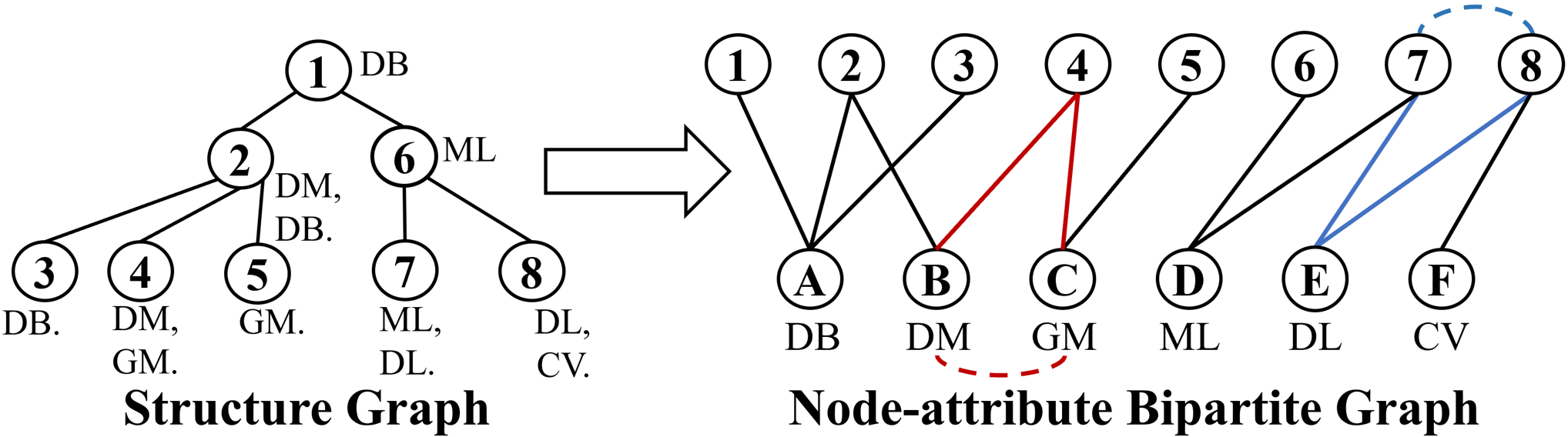}
 %\caption{Illustration for filter of attribute}
 \vspace{-0.4cm}
 \caption{An example of node-attribute bipartite graph.}
 \label{fig:bipart_G_example}
 \vspace{-0.4cm}
\end{figure}

\vspace{-0.2cm}
\subsection{Attribute Encoder}
\vspace{-0.1cm}

The Attribute Encoder provides the interface for query attributes $\mathcal{F}_q$ and produces the vertex embeddings based on the related attributes of queries.  Attribute Encoder aims to figure out the underlying relationship among different attributes and find the related attribute of queries. In addition, as analyzed in Section~\ref{sec.ACS.chanl}, Attribute Encoder needs to represent such attribute information in the form of vertices since the final output community is represented by a set of vertices.%formed by vertices but not by attributes. 

To achieve the above goals, we model a bipartite graph called node-attribute bipartite graph $BG(\mathcal{V}, \hat{\mathcal{F}}, \mathcal{E}_B)$.  For clarity, we call the vertices in the structure graph as nodes here. This bipartite graph is formed by two vertex sets: graph nodes $\mathcal{V}$ and graph attributes $\hat{\mathcal{F}}$. An edge between node $\mathrm{v}_i$ and attribute $\mathrm{f}_j$ is added to the edge set $\mathcal{E}_B$, if and only if node $\mathrm{v}_i$ has attribute $\mathrm{f}_j$, i.e., $\mathrm{f}_j \in \mathcal{F}_i$. 
% Formally, the adjacency matrix of $B$ is defined as: 
% \begin{align}
%  A_{BG}=
% 	\left(
% 	\begin{array}{cc}
% 	0_{V,V} & B_{V} \\ 
% 	B_{F} & 0_{F,F}
% 	\end{array}
% 	\right),
% 	\label{eq:ab}
% \end{align}
% where $B_V = F \in \mathbb{R}^{n \times d}$ and $B_F = F^{\text{T}} \in \mathbb{R}^{d \times n}$.

\vspace{-0.1cm}
\begin{example}
Figure \ref{fig:bipart_G_example} illustrates the node-attribute bipartite graph for the example in Figure~\ref{fig:example1}. Based on the structure graph with node attributes on the left, we construct a node-attribute bipartite graph shown in Figure \ref{fig:bipart_G_example} (right), where the node set $\mathcal{V}=\{1,\dots ,8\}$ is on the top and the attribute set $\hat{\mathcal{F}}=\{A,\dots ,F\}$ is at the bottom. Since node $4$ has two attributes ``DM'' and ``GM'' in the structure graph, node $4$ is adjacent to attribute $B$ (``DM'') and attribute $C$ (``GM'') as connected by red lines in Figure~\ref{fig:bipart_G_example} (right). 
% In Figure~\ref{fig:bipart_G_example}, we can observe that only one vertex ($v_3$) has both attributes $A$ and $B$ while two vertices ($v_3$ and $v_4$) have attributes $B$ and $D$. The relation between $B$ and $D$ is closer than that of $B$ and $A$. 
\end{example}
\vspace{-0.15cm}

We apply Bipartite Graph Neural Network (BGNN) \cite{DBLP:journals/corr/abs-1906-11994} on the constructed bipartite graph.  BGNN consists of propagations in two directions between two vertex sets.  In our node-attribute bipartite graph, the propagations are from the attribute side to the node side (denoted as A$\rightarrow$N), and also from the node side to the attribute side (denoted as  N$\rightarrow$A). 

% \vspace{-0.1cm}
% \stitle{Input Construction}. As introduced in Section~\ref{sec.frame.input}, we encode each query attribute set $\mathcal{F}_q \subseteq \hat{\mathcal{F}}$ to an one hot vector $\boldsymbol{f}_q \in \{0,1\}^{d}$, where $d=|\hat{\mathcal{F}}|$. 
% % For an attribute query $\mathcal{F}_q$, if attribute $\mathrm{f}_i \subseteq \mathcal{F}_q$, value ${\boldsymbol{f}_q}_i=1$, otherwise, ${\boldsymbol{f}_q}_i=0$. 
% For example, when attribute query is $\mathcal{F}_q=\{\text{GNN}\}$, the one-hot vector is $\boldsymbol{f}_q=[0,0,0,0,1,0,0]^T$ according to the order from ``A'' to ``F'' in Figure~\ref{fig:bipart_G_example}. 
\vspace{-0.1cm}
\stitle{Propagation A$\rightarrow$N}. We encode each query attribute set $\mathcal{F}_q \subseteq \hat{\mathcal{F}}$ to a one-hot vector $\boldsymbol{f}_q \in \{0,1\}^{d}$ as described in Section~\ref{sec.frame.input}, where $d=|\hat{\mathcal{F}}|$. 
Benefiting from the node-attribute bipartite graph, we are able to take the query attribute vector $\boldsymbol{f}_q$ as input features in the attribute side, and propagate this attribute information from the attribute side to the node side. 

\vspace{-0.15cm}
% \blue{
\begin{example}
% For example, 
When the attribute query is $\mathcal{F}_q=\{$``DL''$\}$, the one-hot vector is $\boldsymbol{f}_q=[0,0,0,0,1,0]^T$ according to the order of attribute vertices A to F in Figure~\ref{fig:bipart_G_example}. The query attribute information of ``DL'' will propagate to node $7$ and node $8$, the neighbors of ``DL'' vertex, through the blue edges in the bipartite graph of Figure~\ref{fig:bipart_G_example}.
\end{example}
% }
\vspace{-0.15cm}

This propagation from the attribute side to the node side (A$\rightarrow$N) collects attribute features for each node and translates the attribute features to node features. The layer-wise propagation function of A$\rightarrow$N in BGNN is formally defined as:
\vspace{-0.1cm}
\begin{equation}
\begin{aligned}
\boldsymbol{h}_{N_u}^{(l+1)}=\text{SUM}(\{\boldsymbol{h}_{A_\mathrm{f}}^{(l)}  \boldsymbol{W}_{A\rightarrow N}^{(l+1)},\mathrm{f}\in \mathcal{N}_B(\mathrm{u})\}),
% +{{\boldsymbol{h}_{A\rightarrow N}}_v}^{(l)} {{\boldsymbol{W}_{A\rightarrow N}}_{self}}^{(l+1)}
\end{aligned} 
\label{eq:AE-encoder-A->N}	
\vspace{-0.15cm}
\end{equation}
where node $u \in \mathcal{V}$, attribute $\mathrm{f} \in \hat{\mathcal{F}}$, and $\mathcal{N}_B(\mathrm{u})$ is the neighbor set of node $\mathrm{u}$ in the bipartite graph. $\boldsymbol{h}_{N_u}^{(l+1)}\in \mathbb{R}^{ d^{(l+1)}}$ is the hidden feature of node $u$ in the $(l+1)$-th layer, $\boldsymbol{h}_{A_\mathrm{f}}^{(l)}\in \mathbb{R}^{ d^{(l)}}$ is the input feature of attribute $\mathrm{f}$ in the $l$-th layer, and $\boldsymbol{W}_{A\rightarrow N}^{(l+1)}\in \mathbb{R}^{d^{(l)} \times d^{(l+1)}}$ is a learnable parameter matrix in propagation from the attribute side to the node side. 
The input feature of attribute $\mathrm{f}$ in the first layer is equal to the value of attribute $\mathrm{f}$ in the one-hot query attribute vector, i.e.,  $\boldsymbol{h}_{A_\mathrm{f}}^{(0)}={\boldsymbol{f}_q}_\mathrm{f}$.

\vspace{-0.1cm}
\stitle{Propagation N$\rightarrow$A}. 
After the propagation from the attribute side to the node side in the $(l+1)$-th layer, the learned features also need to be transmitted back to form an iterative propagation in the bipartite graph. Here, we also emphasize the attribute in the last layer and add a self feature modeling \cite{self_feat}. Similarly, the layer-wise propagation function from the node side to the attribute side (N$\rightarrow$A) in BGNN is defined as:
\vspace{-0.1cm}
\begin{equation}
\begin{aligned}
\boldsymbol{h}_{A_\mathrm{f}}^{(l+1)}=\boldsymbol{h}_{A_\mathrm{f}}^{(l)} \boldsymbol{W}_\text{self}^{(l+1)}+\text{SUM}(\{\boldsymbol{h}_{N_u}^{(l+1)} \boldsymbol{W}_{N\rightarrow A}^{(l+1)}:\mathrm{\mathrm{u}}\in \mathcal{N}_B(f)\}),
% +{{\boldsymbol{h}_{A\rightarrow N}}_v}^{(l)} {{\boldsymbol{W}_{A\rightarrow N}}_{self}}^{(l+1)}
\end{aligned} 
\label{eq:AE-encoder-N->A}	
\vspace{-0.1cm}
\end{equation}
% where $\boldsymbol{h}_{A_\mathrm{f}}^{(l+1)}\in \mathbb{R}^{ d^{(l+1)}}$ is the learned hidden features for attribute $\mathrm{f}$, $\boldsymbol{h}_{A_\mathrm{f}}^{(l)}\in \mathbb{R}^{ d^{(l)}}$ is the input features of attribute $\mathrm{f}$ in $(l+1)$-th layer. 
where the notations are the same as Eq.~\eqref{eq:AE-encoder-A->N}.  $\boldsymbol{h}_{N_u}^{(l+1)}$ is the input features of node $u$ in propagation N$\rightarrow$A, which is learned in Eq.~\eqref{eq:AE-encoder-A->N}. $\mathcal{N}_B(\mathrm{f})$ is the neighbor set of attribute $\mathrm{f}$ in the bipartite graph. 
$\boldsymbol{W}_{N\rightarrow A}^{(l+1)}\in \mathbb{R}^{d^{(l)} \times d^{(l+1)}}$ is a learnable parameter matrix in the propagation from the node side to the attribute side, and $\boldsymbol{W}_\text{self}^{(l+1)}\in \mathbb{R}^{d^{(l)} \times d^{(l+1)}}$ is the self feature parameter matrix in the $(l+1)$-th layer. 

With these two propagations, Attribute Encoder can employ the query attribute as input features and transmit this attribute information through the node-attribute bipartite graph. Propagation A$\rightarrow$N transforms the attribute features ${\boldsymbol{h}_{A}}$ into node features ${\boldsymbol{h}_{N}}$. 
Propagation N$\rightarrow$A translates the node features ${\boldsymbol{h}_{N}}$ back to attribute features ${\boldsymbol{h}_{A}}$ and provides the input of propagation A$\rightarrow$N in the next layer. 
With these bidirectional propagations, the features can spread in the bipartite graph and BGNN can be superimposed to multiple layers. 
Note that the node features ${\boldsymbol{h}_{N}}$ are the output of Attribute Encoder to Feature Fusion, since the community search problem focuses on the node and other encoders also provide node embeddings rather than attribute embeddings.

\vspace{-0.2cm}
\subsection{Feature Fusion} \label{sec.model.fuse}
\vspace{-0.1cm}
\blue{
The Feature Fusion operator combines the output features of the three encoders, balances the global graph and local query knowledge, and mixes the structure and attribute information to obtain the final output of the \AQD model.
% The Feature Fusion operator combines three output features learned by the Graph Encoder, Query Encoder and Attribute Encoder, which balances the global graph and local query knowledge, and mixes structure and attribute information to obtain the final output of the entire \AQD model. 
% The inputs of Feature Fusion are based on the outputs of three encoders, i.e., ${\boldsymbol{h}_G}_v$, ${\boldsymbol{h}_Q}_v$ and ${\boldsymbol{h}_N}_v$. 
% It fuses them and transmits the fusion result to the Query Encoder and Attribute Encoder as shown in Figure~\ref{fig:framework-b}.
}

The forward layer of Feature Fusion is formulated as:
\vspace{-0.2cm}
\begin{equation}
\begin{aligned}
\boldsymbol{h}_{{FF}_v}^{(l+1)}= \AGG(\boldsymbol{h}_{G_v}^{(l+1)},\boldsymbol{h}_{Q_v}^{(l+1)},\boldsymbol{h}_{N_v}^{(l+1)}),
\end{aligned}
\label{eq:ACS-GCN-Fuse}	
\vspace{-0.25cm}
\end{equation}
where $\boldsymbol{h}_{{FF}_v}^{(l+1)}$ is the fused feature of node $v$ and also the final output of the \AQD model in the $(l+1)$-th layer, $\AGG(\cdot)$ is the aggregation function (e.g., Concatenation, SUM, etc.), and $\boldsymbol{h}_{G_v}^{(l+1)}$, $\boldsymbol{h}_{Q_v}^{(l+1)}$, $\boldsymbol{h}_{N_v}^{(l+1)}$ are the outputs of the three encoders in the $(l+1)$-th layer respectively. Note that $\boldsymbol{h}_{N_v}^{(l+1)}$ is the hidden features of the node side in Attribute Encoder.

In Eq.~\eqref{eq:ACS-GCN-Fuse}, we aggregate the three encoders to fuse all types of node embeddings. In order to consider the correlation between structure and attribute and process these two types of information simultaneously, we replace the input node features in the intermediate layers with the fused feature $\boldsymbol{h}_{{FF}}$ in Query Encoder and Attribute Encoder as shown in Figure~\ref{fig:framework-b}. 
Graph Encoder just uses the output of itself in the $l$-th layer $\boldsymbol{h}_{G}^{(l)}$ as the input of the $(l+1)$-th layer to capture the global query-independent node embeddings, as Feature Fusion does in \QD. 
For Query Encoder, this fusion operation transmits the query-specific attribute features and global graph features into Query Encoder and delivers these features between vertices. Similar to Feature Fusion in \QD, we employ $\hat{\boldsymbol{h}}_Q$ in Eq.~\eqref{eq:hsl} as the input features for Query Encoder, and rewrite the propagation function in Eq.~\eqref{eq:SE-encoder2}.
For Attribute Encoder, Feature Fusion enriches the features passed on the bipartite graph with local query structure and global graph features. Similar to Query Encoder, we replace the input node features in Eq.~\eqref{eq:AE-encoder-N->A} with fused features when propagating from the node side to the attribute side. We define $\hat{\boldsymbol{h}}_N$ as the input node features:
\vspace{-0.15cm}
\begin{equation}
\begin{aligned}
 \hat{\boldsymbol{h}}_{N_u}^{(l)}=\boldsymbol{h}_{{FF}_u}^{(l)}.
 \label{eq:fuse_AE}
\end{aligned}
\vspace{-0.15cm}
\end{equation}
% 示意图解释一下，都是怎么fused的

In this way, the structure features and attribute features learned by \AQD can influence each other and these two encoders are correlated. Thus \AQD is able to learn local structure and related attribute information of queries simultaneously. \AQD provides an end-to-end attributed community search model, which takes queries as input and produces community vectors as answers.

\begin{algorithm}[t]
\footnotesize
\caption{The $k$-Layer \AQD Propagation} \label{algo:AQD}
\begin{flushleft}
\footnotesize
\textbf{Input:} Graph: $G=(\mathcal{V}, \mathcal{E}, \mathcal{F})$, \\
\hspace{0.75cm} a set of attributed queries: $\mathcal{Q}=\{\{{\mathcal{V}_q}_1,{\mathcal{F}_q}_1\},\{{\mathcal{V}_q}_2,{\mathcal{F}_q}_2\},\dots,\}$, \\
\hspace{0.75cm} \AQD model: $\mathcal{M}=\{\boldsymbol{h}_Q, \boldsymbol{h}_G, \{\boldsymbol{h}_N,\boldsymbol{h}_A\}, \boldsymbol{h}_{FF}\}$.\\
\textbf{Output:} a set of output vectors: $\mathcal{H}=\{{{\boldsymbol{h}_q}_1,{\boldsymbol{h}_q}_2}\dots,\}$.\\
\end{flushleft}
%\vspace{0.1cm}
\begin{algorithmic}[1]
\footnotesize
% \STATE  Build \QD $\mathcal{M}$ with two branches SE ($\boldsymbol{h}_Q$) and GE ($\boldsymbol{h}_G$), and a fusion operator ($\boldsymbol{h}_{FF}$).

\STATE  Construct attribute matrix ${\boldsymbol{F}}$ for ${\mathcal{F}}$

\STATE  $\mathcal{H} \leftarrow \varnothing$

\STATE  \textbf{for} each $\{{\mathcal{V}_q},{\mathcal{F}_q}\}\in \mathcal{Q}$ \textbf{do}

\STATE  \hspace{0.3cm} Construct one-hot vector ${\boldsymbol{v}_q}$ for ${\mathcal{V}_q}$, initialize $\boldsymbol{h}_Q^{(0)}$ with ${\boldsymbol{v}_q}$

\STATE  \hspace{0.3cm} Initialize $\boldsymbol{h}_G^{(0)}$ with $\boldsymbol{F}$

\STATE  \hspace{0.3cm} Construct one-hot vector ${\boldsymbol{f}_q}$ for ${\mathcal{F}_q}$, Initialize $\boldsymbol{h}_A^{(0)}$ with ${\boldsymbol{f}_q}$

\STATE  \hspace{0.3cm} $\boldsymbol{h}_Q^{(1)} \leftarrow \text{Propg}(\boldsymbol{h}_{Q}^{(0)}, \mathcal{E})$ in Eq.~\eqref{eq:SE-encoder}%Propagate ${\boldsymbol{h}_{S}}^{(0)}$ through edges in $\mathcal{E}$ and obtain ${\boldsymbol{h}_Q}^{(1)}$ (Eq.~\eqref{eq:SE-encoder2})

\STATE  \hspace{0.3cm} $\boldsymbol{h}_G^{(1)} \leftarrow \text{Propg}(\boldsymbol{h}_{G}^{(0)}, \mathcal{E})$ in Eq.~\eqref{eq:GE-encoder}%Propagate ${\boldsymbol{h}_{G}}^{(0)}$ through edges in $\mathcal{E}$ and obtain ${\boldsymbol{h}_G}^{(1)}$ (Eq.~\eqref{eq:GE-encoder})

\STATE  \hspace{0.3cm}  $\boldsymbol{h}_N^{(1)} \leftarrow \text{Propg}(\boldsymbol{h}_{A}^{(0)}, \mathcal{E}_B)$ in Eq.~\eqref{eq:AE-encoder-A->N}%Propagate ${\boldsymbol{h}_{A}}^{(0)}$ through edges in $\mathcal{E}_B$ and obtain ${\boldsymbol{h}_N}^{(1)}$ (Eq.~\eqref{eq:AE-encoder-A->N})

\STATE  \hspace{0.3cm} $\boldsymbol{h}_{FF}^{(1)} \leftarrow \AGG(\boldsymbol{h}_{G}^{(1)},\boldsymbol{h}_{Q}^{(1)},\boldsymbol{h}_{N}^{(1)})$ in Eq.~\eqref{eq:ACS-GCN-Fuse}

\STATE  \hspace{0.3cm} $l \leftarrow 1$

\STATE  \hspace{0.3cm} \textbf{while}($l<k$) \textbf{do}

\STATE  \hspace{0.3cm} \hspace{0.3cm} $\boldsymbol{h}_Q^{(l+1)} \leftarrow \text{Propg}(\boldsymbol{h}_{FF}^{(l)}, \mathcal{E})$ in Eq.~\eqref{eq:SE-encoder2}%Propagate {${\boldsymbol{h}_{FF}}^{(l)}$} through edges in $\mathcal{E}$ and obtain ${\boldsymbol{h}_Q}^{(l+1)}$ (Eq.~\eqref{eq:SE-encoder2})

\STATE  \hspace{0.3cm}  \hspace{0.3cm} $\boldsymbol{h}_G^{(l+1)} \leftarrow \text{Propg}(\boldsymbol{h}_{G}^{(l)}, \mathcal{E})$ in Eq.~\eqref{eq:GE-encoder}%Propagate ${\boldsymbol{h}_{G}}^{(l)}$ through edges in $\mathcal{E}$ and obtain ${\boldsymbol{h}_G}^{(l+1)}$ (Eq.~\eqref{eq:GE-encoder})

\STATE  \hspace{0.3cm}  \hspace{0.3cm} $\boldsymbol{h}_A^{(l)} \leftarrow \text{Propg}(\boldsymbol{h}_{FF}^{(l)}, \mathcal{E}_B)$ in Eq.~\eqref{eq:AE-encoder-N->A}%Propagate ${\boldsymbol{h}_{N}}^{(l)}$ through edges in $\mathcal{E}_B$ and obtain ${\boldsymbol{h}_A}^{(l)}$ (Eq.~\eqref{eq:AE-encoder-N->A})

\STATE  \hspace{0.3cm}  \hspace{0.3cm} $\boldsymbol{h}_N^{(l+1)} \leftarrow \text{Propg}(\boldsymbol{h}_{A}^{(l)}, \mathcal{E}_B)$ in Eq.~\eqref{eq:AE-encoder-A->N}%Propagate ${\boldsymbol{h}_{A}}^{(l)}$ through edges in $\mathcal{E}_B$ and obtain ${\boldsymbol{h}_N}^{(l+1)}$ (Eq.~\eqref{eq:AE-encoder-A->N})

\STATE  \hspace{0.3cm}  \hspace{0.3cm} $\boldsymbol{h}_{FF}^{(l+1)} \leftarrow \AGG(\boldsymbol{h}_{G}^{(l+1)},\boldsymbol{h}_{Q}^{(l+1)},\boldsymbol{h}_{N}^{(l+1)})$ in Eq.~\eqref{eq:ACS-GCN-Fuse}

\STATE  \hspace{0.3cm} \hspace{0.3cm} $l \leftarrow l+1$

% \STATE  \hspace{0.3cm} Translate ${\boldsymbol{h}_{FF}}^{(3)}$ into node set $\mathcal{C}_q$ with threshold $\gamma$

\STATE  \hspace{0.3cm} $\mathcal{H} \leftarrow \mathcal{H} \cup \boldsymbol{h}_{FF}^{(k)}$

\STATE  \textbf{return} $\mathcal{H}$;

\end{algorithmic}
\end{algorithm}

\vspace{-0.2cm}
\subsection{Algorithm}
\vspace{-0.1cm}

%Based on above three encoder components and a Feature Fusion operator, \AQD model combines one independent graph-level of community information and two correlated query-level information of structure and attribute. \AQD supports all information propagation in the cascades of structures and attributes from the entire graph to the local subgraph of queries. Moreover, \AQD provides an end-to-end multi-query community search model, which takes queries as input and produces community vectors as answers.

Algorithm \ref{algo:AQD} describes the $k$-layer propagation of \AQD.  \AQD first constructs the attribute matrix from $\mathcal{F}$, and builds an empty output set $\mathcal{H}$ (line 1-2). 
For each query, \AQD constructs the one-hot vectors for both query vertex set and query attribute set, and initializes three encoders with them (line 3-6). 
In the first layer (line 7-11), Query Encoder propagates query vertices in the structure graph (line 7), Graph Encoder propagates vertex attributes in the graph (line 8), and Attribute Encoder propagates the query attributes from the attribute side to the node side in the bipartite graph (line 9). Feature Fusion fuses the output features of the three encoders (line 10). 
In the intermediate layers (line 12-18), Query Encoder propagates the fused features in graph (line 13), Graph Encoder still propagates the query-independent features from itself $\boldsymbol{h_G}$ (line 14), and Attribute Encoder utilizes the fused features $\boldsymbol{h}_{FF}$ as node side features and transmits node features back to attribute features (line 15). Then, Attribute Encoder is able to acquire the node hidden features in the next layer through propagating attribute features to the node side in the bipartite graph (line 16).  Feature Fusion fuses the three encoders (line 17).
The final output of \AQD is the fused result in the last layer, which is added into the output set $\mathcal{H}$ (line 19). 

As described in Section~\ref{sec.frame}, in the training stage, the output set $\mathcal{H}$ is used in the loss function to optimize the model learning. In the online query stage, the output is translated to the predicted communities through the Community Identification process as described below.

\vspace{-0.2cm}
\subsection{Community Identification}
\label{Sec.ACS.CI}
 \vspace{-0.1cm}
For the attributed community search problem, we need to find vertices having both dense structure and similar attributes to the query. Thus on top of the online query stage described in Section~\ref{sec.frame.onlineQuery} which ensures connectivity with the query vertices, we also enhance the connectivity between graph vertices sharing identical attributes by a fusion graph $G_F=\{\mathcal{V},\mathcal{E}_F\}$ which combines the information of structure graph $G=\{\mathcal{V},\mathcal{E}\}$ and bipartite graph $G_B=\{\mathcal{V},\hat{\mathcal{F}},\mathcal{E}_B\}$.

%in the online query stage, community identification translates the output vector into community through a threshold $\gamma$ and the constraint that vertices in the community are connected in the graph. The connectivity constraint is only on the structure graph for community search problem which finds the cohesive communities. 
% Accordingly, the constraint can be extended to the connectivity in both structure graph and node-attribute bipartite graph. 
%\blue{Since the final result focuses on the vertices, we build a fusion graph $G_F=\{\mathcal{V},\mathcal{E}_F\}$ with the combine of structure graph $G=\{\mathcal{V},\mathcal{E}\}$ and bipartite graph $G_B=\{\mathcal{V},\hat{\mathcal{F}},\mathcal{E}_B\}$, and input the fusion graph $G_F$ to Algorithm~\ref{algo:BFS}}

\blue{To build the fusion graph, we link vertices with the same attributes in the structure graph. The connectivity in the fusion graph represents both the structure connectivity and attribute similarity. Then the fusion graph $G_F$ is fed to Algorithm~\ref{algo:BFS} for a constrained BFS with the model output $\boldsymbol{h}_q$ for community identification. }

%Then, community identification can achieve the goal to find both structure cohesive and attribute homogeneous communities based on outputs of \AQD. 

%input the fusion graph 

\begin{figure}[t]
 \centering
 \vspace{-0.1cm}
 \includegraphics[width=0.75\linewidth]{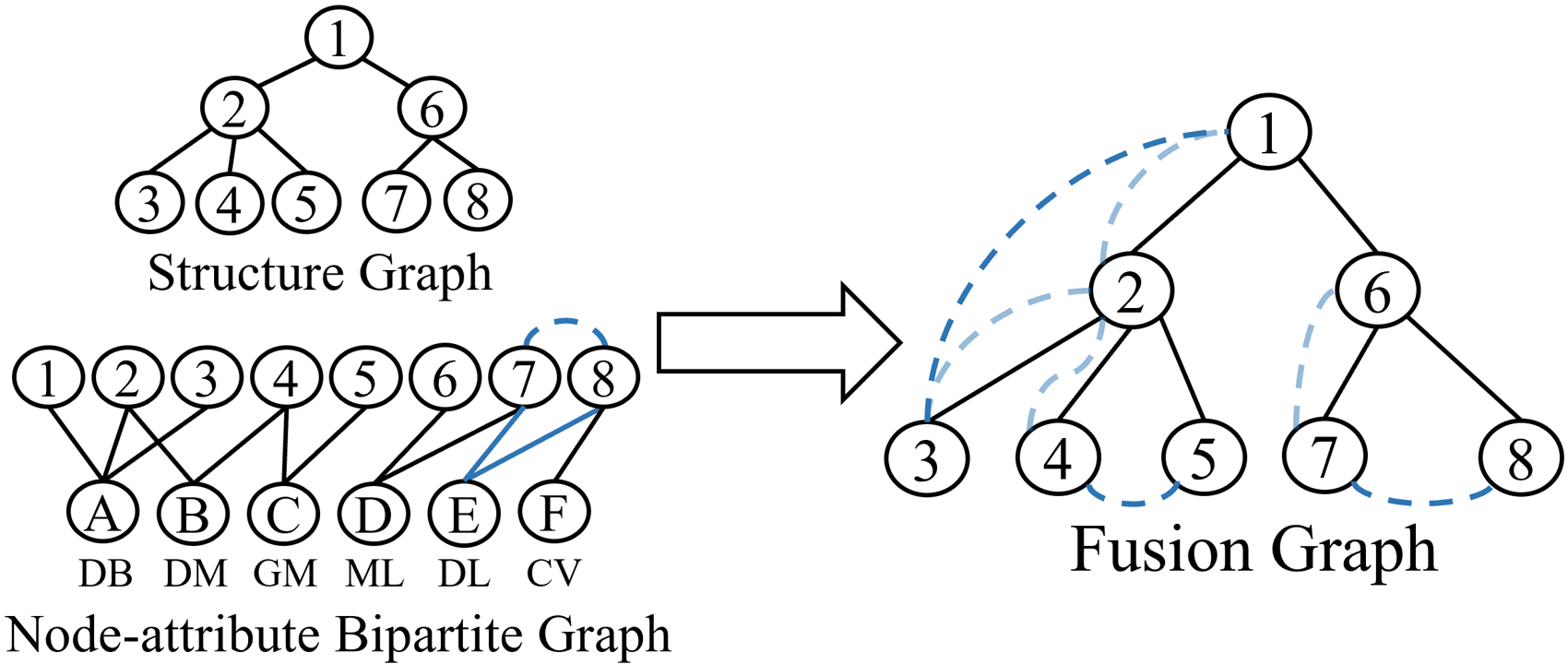}
 %\caption{Illustration for filter of attribute}
 \vspace{-0.4cm}
 \caption{An example of fusion graph.}
 \label{fig:fuseG_example}
 \vspace{-0.3cm}
\end{figure}

\vspace{-0.15cm}
\begin{example}
Figure \ref{fig:fuseG_example} shows the fusion graph for our running example. We add a dashed blue edge between two vertices in the structure graph if they have the same attribute, e.g., vertices 7 and 8 are connected by a dashed blue edge because they both have attribute ``DL''.

%All the edges in $\mathcal{E}$, are also retained in $\mathcal{E}_F$, as the solid black edges of fusion graph in Figure~\ref{fig:fuseG_example}. 
%For bipartite graph, if edge $e_1=(u,f)$ and $e_2=(v,f)$ are both belong to bipartite edge set $\mathcal{E}_B$, then we add a edge $e=(u,v)$ to the fusion edge set $\mathcal{E}_F$. \blue{Demonstrating in Figure~\ref{fig:fuseG_example}, the blue edges $(7,E)$ and $(8,E)$ are in $\mathcal{E}_B$, and we add an edge $(7,8)$ in the fusion graph marked as blue dash line.}
\end{example}
\vspace{-0.15cm}

\vspace{-0.3cm}
\subsection{Complexity Analysis}
\label{sec.ACS.complx}
\vspace{-0.1cm}

\blue{In order to analyze the time complexity of \QD and \AQD,} we first present the complexity of general GNN in Eq.\eqref{eq:generalGCN}. This GNN aggregates neighbors' features for every vertex with the cost of $\sum_{i=\mathrm{u}}^n{d_\mathrm{u}}$, where $d_\mathrm{u}$ is the degree of vertex $\mathrm{u}$ and $n$ is number of vertices. Thus the complexity of general GNN is $O(|\mathcal{E}|)$.

\blue{For \QD, Query Encoder and Graph Encoder have the same time complexity of $O(|\mathcal{E}|)$ as general GNN.} For \AQD, the time cost of Attribute Encoder is also dependent on the sum of vertices' degree in the bipartite graph with the complexity of $O(|\mathcal{E}_B|)$. The aggregation operation in Feature Fusion, e.g., MAX, Concatenation, etc., is implemented in parallel and the complexity is just $O(1)$. 
\blue{Suppose \QD or \AQD is a $k$-layer model with $t$ iterations, where $k=3$ and $t=300$ are typical settings. The complexity of \QD is $O(k\times t \times|\mathcal{E}|)$ in the model training stage and $O(k\times|\mathcal{E}|)$ in the online query stage. 
Similarly, the complexity of \AQD is $O(k\times t \times(|\mathcal{E}|+|\mathcal{E}_B|))$ in the model training stage and $O(k\times(|\mathcal{E}|+|\mathcal{E}_B|))$ in the online query stage.}

% QE: $O(|\mathcal{E}|)$ or *d
% GE: $O(|\mathcal{E}|)$
% AE: $O(|\mathcal{E}_B| )$
% Fusion: $O(d)$
% CI: $O(|\mathcal{E}|+|\mathcal{E}_B|)$

% \red{Here}

% \vspace{-0.2cm}
\section{EXPERIMENTS}\label{sec.exp}
\vspace{-0.1cm}

In this section, we present our experimental studies to validate the performance of 
%three proposed models on real graphs: \SQD, \QD and \AQD.
our framework with the three proposed models in different scenarios. 
We first introduce the setup of our experiment in Section \ref{sec.setup}. Then we evaluate the performance in both attributed and non-attributed community search problem in Section \ref{sec.result}. To further verify the effectiveness of our models, we compare our models with the interactive community search model ICS-GNN in Section~\ref{sec.exp.ics}. 
% To evaluate the efficiency of 
%\AQD
% our model \AQD for the attributed community search, we also test the response time of the queries in Section~\ref{sec.efficiency}. 
\blue{
Moreover, we evaluate the performance of our model for ACS on large graphs in Section~\ref{sec.exp.large}. 
Finally, we conduct the ablation study in Section~\ref{sec.ablation} to demonstrate the effectiveness of Feature Fusion, the sensitivity test of the parameter $\gamma$, the data split ratio, the epoch number and dropout rate. 
}
%Finally, we perform case studies on two data sets in Section~\ref{sec.case} to show its representation power in real applications.  

\vspace{-0.2cm}
\subsection{Experimental Setup} \label{sec.setup}
\vspace{-0.05cm}

\renewcommand\arraystretch{0.9}
\begin{table}
 \footnotesize
    \centering
    \caption{Dataset Statistics. $|\mathcal{V}|$ and $|\mathcal{E}|$ are the number of vertices and edges. $|\hat{\mathcal{F}}|$ is the number of distinct attributes, $K$ is the number of communities and $AS$ is the average size of communities. Here, \textbf{M}$=10^6$.}%{ $|\mathcal{V}|$ is the number of vertices, $|\mathcal{E}|$ is the number of edges, $|\mathcal{F}|$ is the number of distinct attributes, $K$ is the number of communities and $AS$ is the average size of communities.}}
    \vspace{-0.4cm}
    % \setlength{\tabcolsep}{1.0mm}
    % \begin{threeparttable*}    %\setlength{\tabcolsep}{0.012\textwidth}{%
    \begin{tabular}{c|l||c|c|c|c|c|c}
    \toprule
        \multicolumn{2}{c||}{ Data set} & $|\mathcal{V}|$ & $|\mathcal{E}|$ & $|\hat{\mathcal{F}}|$ & $|\mathcal{E}_B|$ & $K$ & $AS$ \\
        \hline
        \multirow{5}*{  } &  {Cornell} & 195 & 283 & 1703 & 18496 & 5 & 39\\
         \cline{2-8}
         &  {Texas} & 187 & 280 & 1703 & 15437 & 5 & 37.4\\
         \cline{2-8}
         Citation  &  {Washt} & 230 & 366 & 1703 & 19953 & 5 & 46\\
         \cline{2-8}
         Networks&  {Wiscs} & 265 & 459 & 1703 & 25479 & 5 & 53\\
        \cline{2-8}
         &  {Cora} & 2708 & 5278 & 1433 & 49216 & 7 & 386.86\\
         \cline{2-8}
         &  {Citeseer} & 3312 & 4536 & 3703 & 105165 & 6 & 552\\
        \hline
         \multirow{9}*{ }&  {Reddit} & 232965 & 114\textbf{M} & 602 & 140\textbf{M} & 50 & 4659.3\\
 %        \hline
 %        {Philosophers} & 1218 & 5972 & 5770 & 1220 & 10.97\\
 %        \hline
 %        {Flicker} & 16710 & 176063 & 1156 & 5436 & 273.10\\
        \cline{2-8}
        &  {FB-0} & 348 & 2852 & 224 & 3348 & 24 & 13.54\\
         \cline{2-8}
          &  {FB-107} & 1046 & 27783 & 576 & 11827 & 9 & 55.67\\
         \cline{2-8}
         Social &  {FB-1684} & 793 & 14810 & 319 & 6131 & 17 & 45.71\\
        \cline{2-8}
         Networks &  {FB-1912} & 756 & 30772 & 480 & 8066 & 46 & 23.15\\
         \cline{2-8}
          &  {FB-3437}& 548 & 5347 & 262 & 4263 & 32 & 6\\
         \cline{2-8}
         &   {FB-348}& 228 & 3416 & 161 & 2398 & 14 & 40.5\\
         \cline{2-8}
          &  {FB-414} & 160 & 1843 & 105 & 1566 & 7 & 25.43\\
         \cline{2-8}
         &  {FB-686}& 171 & 1824 & 63 & 999 & 14 & 34.64\\
    \toprule
    \end{tabular}%}
    \label{tab:statis}
    \vspace{-0.5cm}
\end{table}

\subsubsection{Data Sets}

To thoroughly evaluate the performance of %three proposed models,
our framework, we conduct experimental studies on 15 attributed graphs. 
%The graph statistics are presented in Table \ref{tab:statis}. 
Table~\ref{tab:statis} reports the dataset statistics.
The first six networks, Cornell, Texas, Washington (Washt), Wisconsin (Wiscs), Cora and Citeseer, are publication citation networks.  Each attribute describes the absence/presence of one word in a publication. All these graphs can be found at LINQS website\footnote{https://linqs.soe.ucsc.edu/data}. Reddit \cite{hamilton2017inductive} is an online discussion website where each vertex is a post and an edge links two posts if they have comments from the same user.
%{\color{red} Pilosophers can be found in \cite{ahn2010link}. can not find the describe of this data.} And Flickr \cite{ruan2013efficient} is a social network where vertices are users and edges are the contacts between users. The attributes of users are the tags of photos user uploaded. And the ground-truth communities are Flickr user groups.
Facebook \cite{McAuleyL12} is a social network where vertices are users and edges are friend relationships. It contains 8 ego-networks with different attributes as shown in Table \ref{tab:statis}. 
We consider each ego-network as an independent data set. 
All data sets contain ground-truth communities. 
%graph and circles in the ego-network as the ground-truth communities.

\vspace{-0.15cm}
\subsubsection{Baseline Models}

We compare our models with five state-of-the-art approaches, including two non-attributed community search algorithms: \CTC \cite{huang2015approximate} and $k$-\ECC \cite{chang2015index}, two attributed community search algorithms: \ACQ \cite{fang2016effective} and \ATC \cite{huang2017attribute}, and a GNN-based interactive community search model ICS-GNN \cite{gao2021ics_ICS-GNN}. % as follows.
% The detailed explanations of each compared methods are in \red{cite}.

\vspace{-0.15cm}
%\subsubsection{Query Generation} 
\subsubsection{Query Setting}
For each data set, we generate $n_q=350$ pairs of input query set $\mathcal{Q}=\{<\mathcal{V}_q, \mathcal{F}_q>\}_{q=1}^{n_q}$ and the corresponding ground-truth community ${\mathcal{Y}}_q$. 
To generate the query vertex set $\mathcal{V}_q$, vertex sets containing 1-3 vertices are randomly selected from the ground-truth community. To generate the query attribute set $\mathcal{F}_q$, we design three different types as described below for fair comparison with different existing CS and ACS methods. The query vertex set and corresponding ground-truth communities are shared across the three types of input queries. 

%in experiments, that is, query with empty attribute (EmA), query with attribute from communities (AFC) and attribute from query nodes (AFN). 
\begin{itemize}[leftmargin=*]
\vspace{-0.1cm}
\item \textbf{ Empty attribute query (EmA)}. To compare with methods for non-attributed community search, we set the attribute query set empty ($\mathcal{F}_q = \emptyset$) and generate the EmA set $\mathcal{Q}_{\mathrm{EmA}} = \{<\mathcal{V}_q, \emptyset>\}$.

\vspace{-0.05cm}
\item \textbf{Attribute from community (AFC)}. As suggested by \cite{fang2016effective,huang2017attribute}, to construct the query attribute set ($\mathcal{F}_q= \mathcal{F}_{q}^{\mathrm{c}}$), we use 5 most common attributes in ground-truth communities. Therefore, we have $\mathcal{Q}_{\mathrm{AFC}} = \{<\mathcal{V}_q, \mathcal{F}_{q}^{\mathrm{c}}>\}$. AFC is used to validate the contribution of the attributes in the community search.

\vspace{-0.05cm}
\item \textbf{Attribute from node (AFN)}. %Furthermore, 
We simulate real queries provided by users and select 5 most common attributes from attributes of query vertices as the query attribute set, i.e,  $\mathcal{F}_q= \mathcal{F}_{q}^{\mathrm{n}}$. In other words, $\mathcal{F}_{q}^{\mathrm{n}}$ may be unrelated to the ground-truth communities. We construct the AFN set $\mathcal{Q}_{\mathrm{AFN}} = \{<\mathcal{V}_q, \mathcal{F}_{q}^{\mathrm{n}}>\}$.
Obviously, AFN is a more challenging setting and closer to the real scenarios. 

\end{itemize}
\vspace{-0.1cm}

\vspace{-0.15cm}
\subsubsection{Data Splitting} For each data set, we split 350 query-community pairs into training data, validation data and test data with the ratio of 150:100:100 by default. We use training data to train our models, validation data to select the best weights during the training process, and test data to measure the performance of all methods. \blue{In the ablation study, we vary the data splitting ratio to evaluate its influence on the performance.}

\vspace{-0.15cm}
\subsubsection{Evaluation Metrics} Let $D_{\text{test}}=\{\mathcal{Q},\hat{\mathcal{C}},\mathcal{Y}\}$ be the test data set, where $\mathcal{Q}$ is the query set, $\hat{\mathcal{C}}$ is the predicted community set by a method and $\mathcal{Y}$ is the ground-truth community set. To measure the quality of communities found by different methods, we employ 
% two measurements:  
F1-score 
% and Jaccard similarity 
to evaluate the quality of the predicted set $\hat{\mathcal{C}}$. 
% \begin{itemize}
    % \item 
    F1-score is defined as:
    \vspace{-0.2cm}
    \begin{equation}
    \nonumber
    \begin{aligned}
        F1(\hat{\mathcal{C}},\mathcal{Y})=\frac{2\cdot pre(\hat{\mathcal{C}},\mathcal{Y}) \cdot rec(\hat{\mathcal{C}},\mathcal{Y})}{pre(\hat{\mathcal{C}},\mathcal{Y}) + rec(\hat{\mathcal{C}},\mathcal{Y})}
    \end{aligned}
    \vspace{-0.2cm}
    \end{equation}
    % $F_1(Z,Y)=\frac{2\cdot pre(Z,Y) \cdot rec(Z,Y)}{pre(Z,Y) + rec(Z,Y)}$. 
    where $pre(\hat{\mathcal{C}},\mathcal{Y})$ is the precision of predicted community set $\hat{\mathcal{C}}$ on the ground-truth community set $\mathcal{Y}$, $rec(\hat{\mathcal{C}},\mathcal{Y})$ is the recall of the predicted communities:
    \vspace{-0.2cm}
     \begin{equation}
    \begin{aligned}
    \nonumber
        pre(\hat{\mathcal{C}},\mathcal{Y})=\frac{\sum_{\boldsymbol{c}_q \in \hat{\mathcal{C}}} \boldsymbol{c}_q \& \boldsymbol{y}_q}{\sum_{\boldsymbol{c}_q \in \hat{\mathcal{C}}} \sum_{i=0}^n \boldsymbol{c}_{qi}}, 
        rec(\hat{\mathcal{C}},\mathcal{Y})=\frac{\sum_{\boldsymbol{c}_q \in \hat{\mathcal{C}}} \boldsymbol{c}_q \& \boldsymbol{y}_q}{\sum_{\boldsymbol{y}_q\in \mathcal{Y}} \sum_{i=0}^n \boldsymbol{y}_{qi}}.
    \end{aligned}
    \vspace{-0.13cm}
    \end{equation}
    % where Precision $pre(P,Y)=\frac{\sum_{z_q \in Z} z_q \& y_q}{\sum_{z_q \in Z} \sum_{i=0}^n z_{qi}}$ and recall $rec(Z,Y)=\frac{\sum_{z_q \in Z} z_q \& y_q}{\sum_{y_q\in Y} \sum_{i=0}^n y_{qi}}$.
    % \item  Jaccard similarity is defined as:
    % \begin{equation}
    % \nonumber
    %  Jac(\hat{\mathcal{C}},\mathcal{Y})=\frac{\sum_{\boldsymbol{c}_q \in \hat{\mathcal{C}}} z_q \& y_q}{\sum_{\boldsymbol{c}_q \in \hat{\mathcal{C}}} \boldsymbol{c}_q | y_q}. 
    %  \end{equation}
% \end{itemize}
Here, $\boldsymbol{c}_q\in {\{0,1\}}^{n\times 1}$ and $y_q\in {\{0,1\}}^{n\times 1}$ are the predicted and ground-truth community vectors for query $q$ respectively.

\vspace{-0.15cm}
\subsubsection{Implementation Details}
\label{sec.exp.setting}
In our models, we build three layers with 128 neurons in the hidden layer. We train 300 iterations with a learning rate of 0.001. 
% For the final output $z_q$, we apply a threshold $\gamma \in (0,1)$ to filter out the output community set $\hat{\mathcal{C}}_q$. If $z_{qi} \geq \gamma$, $v_i$ belongs to the output community, then $\hat{z}_{qi}=1$. We choose $\gamma$ to a value that achieves the best performance on the validation query set. 
In the Feature Fusion component, we choose concatenate as the aggregation function in Eq.~\eqref{eq:FuseCS-GCN} and Eq.~\eqref{eq:ACS-GCN-Fuse}.
\blue{In each layer except the output layer, we employ ReLU as activation function, batch normalization with batch size $4$ and dropout rate $0.5$ \cite{srivastava2014dropout} for each branch.}

\vspace{-0.2cm}
\subsection{Community Search Performance} \label{sec.result}
\vspace{-0.1cm}

We present comprehensive experiments to validate the query performance of the three proposed models under two settings: non-attributed community search, and attributed community search. 
% In addition, we also analyze the efficiency for this online query problem.
%test score/training time  figure to show the training time compare with others index time

\vspace{-0.1cm}
\begin{figure}
\small
    \centering
    \includegraphics[width=1.0\linewidth]{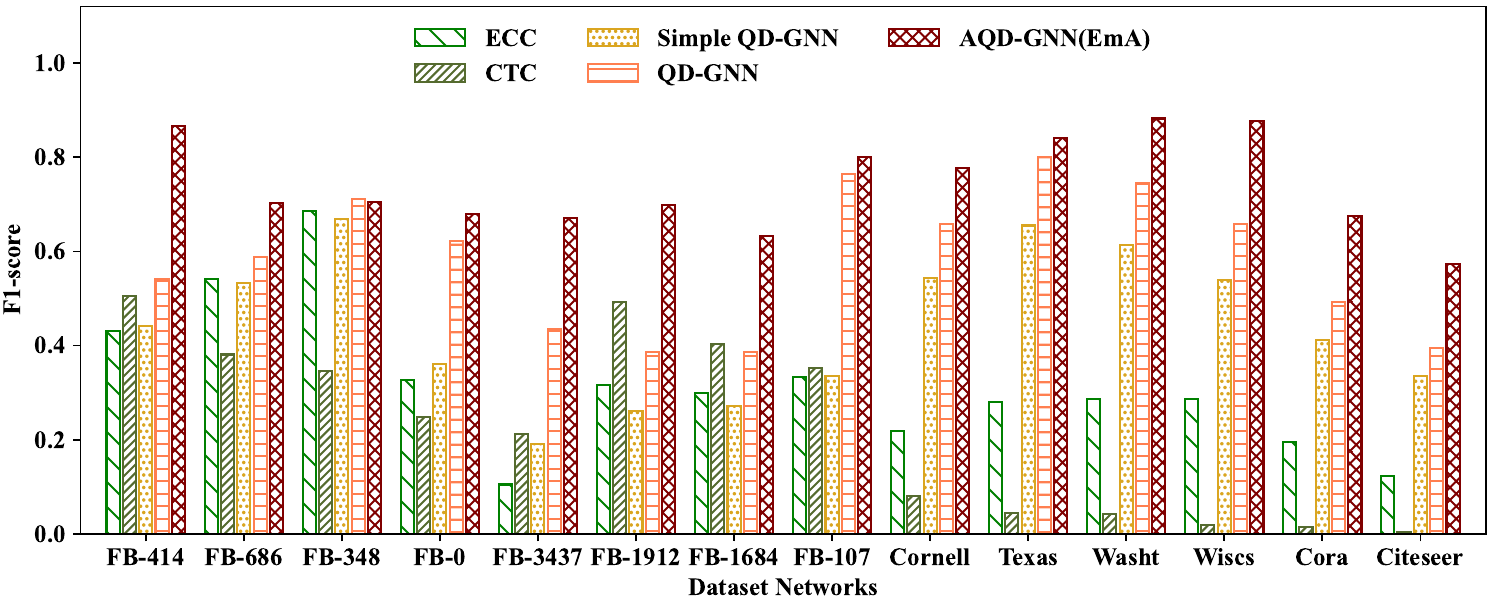}
    \vspace{-0.8cm}
    \caption{Non-attributed community search performance comparison.}
    \label{fig:result-NonAtt}
    \vspace{-0.5cm}
\end{figure}

\begin{figure*}[t]
% 	\vspace{-0.1in}
\vspace{-0.6cm}
	\centering
	%  % \hspace{-0.04\textwidth}
% 	\hspace{-0.2cm}
	\begin{subfigure}[b]{0.49\textwidth}
		\centering
		\includegraphics[width=1.0\textwidth]{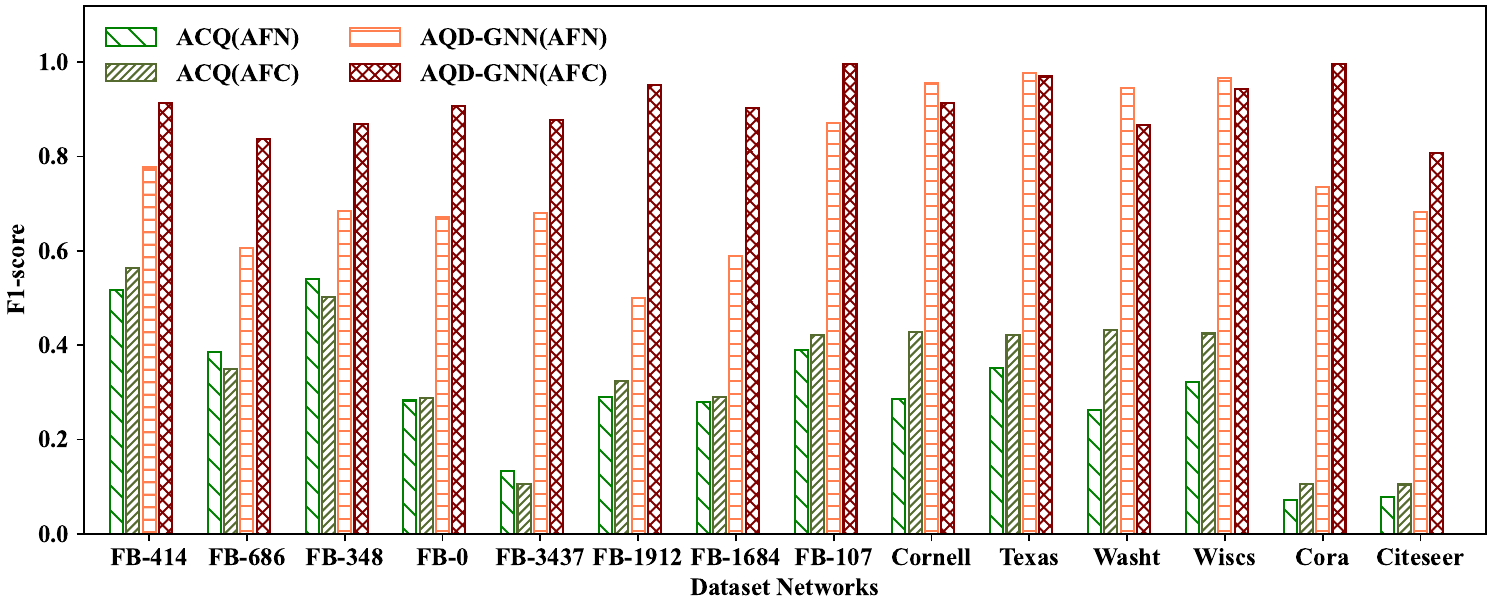}
		\vspace{-0.6cm}
		\caption{Compared with methods supporting one-vertex queries.}
		\label{fig:result-ACQ}
	\end{subfigure}
	\begin{subfigure}[b]{0.49\textwidth}
		\centering
% 		\hspace{-0.1cm}
% 		\includegraphics[width=1.02\textwidth]{./figure/exp/CS_ATC.pdf}
% 		\includegraphics[width=1.0\textwidth]{./figure/ATC.pdf}
		\includegraphics[width=1.0\textwidth]{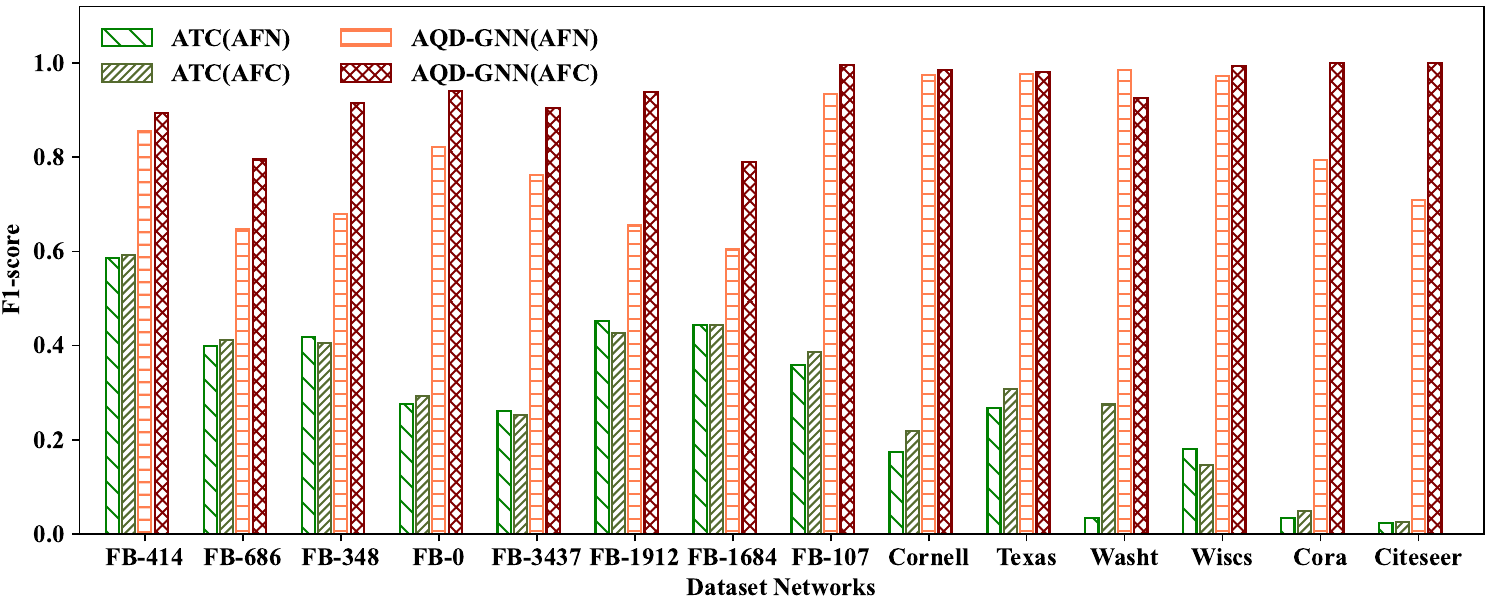}
		\vspace{-0.6cm}
		\caption{Compared with methods supporting multi-vertex queries.}
		\label{fig:result-ATC}
	\end{subfigure}
	\vspace{-0.5CM}
	\caption{Attributed community search performance compared with other approaches.}
	\label{fig:result-ACS}
	\vspace{-0.1CM}
\end{figure*}

%${H_F}^{(0)}=[0,0, \dots, 0]^T$
\vspace{-0.15cm}
\subsubsection{Non-attributed community search} \label{sec.noa_res}
In order to compare to non-attributed community search algorithms, we generate the multi-vertex queries set without query attributes $\mathcal{Q}_{\mathrm{EmA}}$, and compare our three models \SQD, \QD, \AQD with \CTC and $k$-\ECC. Figure \ref{fig:result-NonAtt} shows the F1-score. We can observe that:
\begin{itemize}[leftmargin=*]
\vspace{-0.1cm}
\item \CTC performs reasonably well in Facebook ego networks but poorly in citation networks, since \CTC searches communities using the $k$-truss subgraph pattern, which may fit the dense social networks well, but does not fit the sparser citation networks.
%relies on $k$-truss as the pre-defined subgraph pattern which may not fit for the citation networks.
%the structure inflexibility of pre-defined subgraph pattern which cannot fit into any graphs.

\vspace{-0.01cm}
\item By capturing the local query structure only, \SQD can outperform \ECC and \CTC in citation networks and achieve comparable performance in Facebook ego networks. It demonstrates the learning-based models can apply to different types of networks and discover communities with different structural properties.

%the flexibility of the learning-based model in the structural properties of. 
%Benefiting from learning based model, \SQD performs more stable, which achieves a comparable performances in Facebook ego network with \ECC and \CTC, and outperforms them in citation networks. 
\vspace{-0.03cm}
\item \QD can substantially outperform \SQD by improving the F1-score by $0.14$ on average. It validates the effectiveness of the query-independent graph features learned from Graph Encoder. 

%Within global graph information, improved model \QD significantly outperforms \SQD among all data sets and increases $0.14$ in average of F1-score. It validates the effectiveness of
%It demonstrate that the query-independent features learning from Graph Encoder can greatly improve the performance of the community search problem.  

\vspace{-0.05cm}
\item We also apply \AQD to non-attributed community search, where we set the query attribute set to empty, $\mathcal{F}_q=\emptyset$. Interestingly, \AQD can achieve the best performance in almost all data sets in Figure~\ref{fig:result-NonAtt}. This is owing to the Feature Fusion operator and Attribute Encoder design in \AQD. Specifically, Feature Fusion can transmit graph information and query vertices information to Attribute Encoder before the second layer. Then Attribute Encoder can utilize the information from the second layer and learn hidden relations between attributes. 

\end{itemize}
\vspace{-0.05cm}

\vspace{-0.15cm}
\subsubsection{Attributed community search}
\label{sec.exp.AQD}
We compare \AQD with two attributed community search algorithms: \ACQ and \ATC. \ACQ can only handle one query vertex while \ATC and our model \AQD can handle multiple query vertices. Thus we compare \ACQ and \AQD for one-vertex queries in Figure~\ref{fig:result-ACQ}, and compare \ATC and \AQD for multi-vertex queries in Figure~\ref{fig:result-ATC}. We can observe that: 
\begin{itemize}[leftmargin=*]
\vspace{-0.15cm}
    \item For Cora and Citeseer with large ground-truth communities (hundreds of vertices in a community), the performances of \ATC and \ACQ are quite poor (around $0.1$ in F1-score). It is because their pre-defined community patterns (i.e., $k$-core and $k$-truss) are too strict to find large communities in the real-world graphs. 
    \vspace{-0.25cm}
    \item \AQD consistently performs the best on all data sets. 
    % Even with empty query attributes (EmA), \AQD still achieves least $0.3$ improvements against \ATC and \ACQ. 
    As a data-driven approach, \AQD is capable of learning communities with varied sizes and shapes. It performs stably on all graphs benefiting from learning adaptive weight matrices for different data sets.
    % \vspace{-0.05cm}
    % \item The performance of AFC and AFN is better than that of EmA in most datasets. It indicates that introducing query attribute information can improve the performance of community search.  
    \vspace{-0.05cm}
    \item Compared to AFC, all methods suffer from performance degradation under the AFN setting in most data sets, since AFC is a more favorable setting where the query attribute set is directly extracted from the most common attributes of the ground-truth. For example, in the Washington data set in Figure~\ref{fig:result-ATC}, \ATC achieves 0.275 F1-score under AFC, but only 0.033 under AFN. 
    % \blue{It is still possible that methods under AFN perform better than under AFC, since AFN may contain more attributes and include general attributes as Section~7.1.3 explains.}
    \vspace{-0.05cm}
    \item Under the more realistic but challenging AFN setting, we can observe \AQD achieves a significant performance improvement over the baselines, with $0.46$ and $0.53$ improvements on F1-score for one-vertex queries and multi-vertex queries respectively. This is because \AQD exploits the node-attribute bipartite graph to find similar attributes, while the baselines simply require vertices in a community have identical attributes with query attributes. 
\end{itemize}
\vspace{-0.15cm}

% ********************************query time*****************************

\renewcommand\arraystretch{0.8}
\begin{table*}
\footnotesize
    \centering
    \vspace{-0.2cm}
    \caption{Average query time (in milliseconds) of different community search methods.}%{ $|\mathcal{V}|$ is the number of vertices, $|\mathcal{E}|$ is the number of edges, $|\mathcal{F}|$ is the number of distinct attributes, $K$ is the number of communities and $AS$ is the average size of communities.}}
    \vspace{-0.4cm}
    \setlength{\tabcolsep}{0.6mm}
    \begin{tabular}{lc||r|r|r|r|r|r|r|r|r|r|r|r|r|r|r}
    \toprule
        \multicolumn{2}{c||}{ Methods} & FB-414 & FB-686 & FB-348 & FB-0 & FB-3437 & FB-1912 & FB-1684 & FB-107 & Cornell & Texas & Washt & Wiscs & Cora & Citeseer& Average  \\
        % \hline
        \midrule
        \multirow{2}*{}Non-&\CTC & 34.78 & 41.41 & 120.92 & 49.25 & 131.67 & 4903.38 & 604.67 & 2498.82 & 0.45 & 0.40 & 0.42 & 0.63 & 1.96 & 1.28 & 599.00 \\
        % \hline
       Attributed &\ECC & 3.52 & 2.29 & 5.20 & 2.57 & 6.60 & 154.23 & 26.85 & 93.62 & 0.24 & 0.28 & 0.23 & 0.33 & 2.67 & 1.76 & 21.50 \\
        \hline
        \specialrule{0em}{0.15mm}{0.1mm}
        \multirow{3}*{}&\ACQ & <0.01 & <0.01 & 1.45 & <0.01 & <0.01 & <0.01 & <0.01 & <0.01 & <0.01 & <0.01 & <0.01 & <0.01 & <0.01 & <0.01 & 0.10 \\
        % \hline
        Attributed&\ATC & 4.40 & 5.10 & 7.70 & 5.90 & 10.30 & 43.60 & 22.70 & 40.10 & 7.90 & 14.02 & 2.11 & 18.60 & 11.49 & 3.68 & 9.80 \\
        % \hline
        % &\AQD & 4.08 & 3.40 & 5.04 & 3.70 & 4.48 & 4.46 & 5.60 & 5.44 & 4.45 & 4.49 & 5.50 & 5.23 & 5.54 & 5.25 & 4.56 \\ %old?
        &\AQD & 3.31 & 3.32 & 3.41 & 3.63 & 4.32 & 4.96 & 4.56 & 5.46 & 4.15 & 4.10 & 4.16 & 4.41 & 5.54 & 5.32 & 4.31 \\ %
        \specialrule{0em}{0.0mm}{0.0mm} 
    \bottomrule
    \end{tabular}%}
    \label{tab:query-time}
    \vspace{-0.1cm}
\end{table*}

% \begin{figure}
% \small
%     \centering
%     \includegraphics[width=0.9\linewidth]{./figure/exp/time_ICS.pdf}
%     \vspace{-0.5cm}
%     \caption{Time comparison with ICS-GNN.}
%     \label{fig:ICS-time}
%     \vspace{-0.6cm}
% \end{figure}  

\vspace{-0.15cm}
\subsubsection{Query Efficiency}
\label{sec.efficiency}

We evaluate the query efficiency of \AQD in the test set.  Table \ref{tab:query-time} shows the average query time (in milliseconds) of 100 test queries by \AQD and baselines. The last column reports the average query time among all data sets. 

Overall, the query time \AQD is much faster than that of all baselines except \ACQ.  \ACQ is a simple baseline which only allows one query vertex and considers vertices' degrees and common attributes.  Its query performance in terms of F1-score is very poor as shown in Figure \ref{fig:result-ACS}.  It is worth noting that \AQD achieves a stable query time of around 5 milliseconds on all data sets, while the query time of \CTC, \ECC and \ATC increases significantly when the graph is large. In particular, \CTC takes almost $5,000$ milliseconds for a query on FB-1912, while \AQD only costs $4.96$ milliseconds. This experiment shows that \AQD is more suitable for online search in real-world applications.

\renewcommand\arraystretch{0.9}
\begin{table*}
\footnotesize
\vspace{-0.15cm}
    \centering
    \caption{F1-score (in $\%$) and time cost (in seconds) of interactive community search methods on different networks.}
    \vspace{-0.4cm}
    \setlength{\tabcolsep}{0.4mm}
    \begin{tabular}{c||cc|cc|cc|cc|cc|cc|cc|cc|cc|cc|cc|cc}
    \toprule
        Method & \multicolumn{2}{c|}{FB-414} & \multicolumn{2}{c|}{FB-686} & \multicolumn{2}{c|}{FB-348} & \multicolumn{2}{c|}{FB-0} & \multicolumn{2}{c|}{FB-3437} & \multicolumn{2}{c|}{FB-1912} & \multicolumn{2}{c|}{FB-1684} & \multicolumn{2}{c|}{FB-107} & \multicolumn{2}{c|}{Cora} & \multicolumn{2}{c|}{Citeseer} & \multicolumn{2}{c|}{Reddit} & \multicolumn{2}{c}{Average}  \\
        % \hline
         & F1 & Time & F1 & Time & F1 & Time & F1 & Time & F1 & Time & F1 & Time & F1 & Time & F1 & Time & F1 & Time & F1 & Time & F1 & Time & F1 & Time\\
        \midrule
        ICS-GNN & 56.63 & 0.14 & 43.53 &  0.15 & 33.88 &  0.26 & 24.94 &  0.26 & 26.49 & 0.51 &  20.23 & 2.36  & 20.76 & 1.28  & 36.46  & 2.40  & 30.52 &  0.14  & 30.29  & 0.14 & 19.41 & 1.84 & 31.19  &  0.86\\
        % \hline
        % \QD & \bf{62.02}  & 0.14 & \bf{44.28} & 0.14  & 34.74 & 0.26  & 31.07 & 0.27  & 27.38 & 0.52  & 21.10 & 2.45  & 24.79 & 1.31  & 38.39  & 2.49  & 32.56 & 0.12  & 31.53  & 0.12 & 21.29  & 1.68 & $33.56_{+2.37}$ & $0.87_{+0.01}$  \\
        % \QD & \bf{62.02}  & 0.14 & \bf{44.28} & 0.14  & 34.74 & 0.26  & 31.07 & 0.27  & 27.38 & 0.52  & 21.10 & 2.45  & 24.79 & 1.31  & 38.39  & 2.49  & 32.56 & 0.12  & 31.53  & 0.12 & 21.29  & 1.68 & 33.56 & 0.87  \\
        \QD & \textbf{62.02}  & 0.14 & \textbf{44.28} & 0.14  & 34.74 & 0.26  & 31.07 & 0.27  & 27.38 & 0.52  & 21.10 & 2.45  & 24.79 & 1.31  & 38.39  & 2.49  & 32.56 & 0.12  & 31.53  & 0.12 & 21.29  & 1.68 & $33.56_{+2.37}$ & $0.87_{+0.01}$  \\
        % \hline
        % \AQD & 57.34  & 0.14 & 38.87  & 0.14 & \bf{35.35} & 0.25  & \bf{35.63}  & 0.26 & \bf{30.22} & 0.50  & \bf{37.52} & 2.29  & \bf{37.91} & 1.23  & \bf{49.67}  & 2.41  & \bf{33.19} & 0.14  & \bf{31.77} & 0.12  & \bf{24.86} & 1.85  & $\bf{37.48}_{+6.29}$  & $0.85_{-0.01}$ \\
        % \AQD & 57.34  & 0.14 & 38.87  & 0.14 & \bf{35.35} & 0.25  & \bf{35.63}  & 0.26 & \bf{30.22} & 0.50  & \bf{37.52} & 2.29  & \bf{37.91} & 1.23  & \bf{49.67}  & 2.41  & \bf{33.19} & 0.14  & \bf{31.77} & 0.12  & \bf{24.86} & 1.85  & \bf{37.48}  & 0.85 \\
        \blue{AQD (AFN)} & \blue{61.25}  & \blue{0.14} & \blue{43.05}  & \blue{0.14} & \blue{\textbf{35.80}} & \blue{0.25}  & \blue{31.27}  & \blue{0.26} & \blue{29.03} & \blue{0.50}  & \blue{20.44} & \blue{2.29}  & \blue{36.51} & \blue{1.23}  & \blue{41.07}  & \blue{2.41}  & \blue{31.81} & \blue{0.14}  & \blue{\textbf{33.09}} & \blue{0.12}  & \blue{20.71} & \blue{1.85}  & \blue{${{34.91}}_{+3.72}$}  & \blue{$0.85_{-0.01}$}  \\
        AQD (AFC) & 57.34  & 0.14 & 38.87  & 0.14 & {35.35} & 0.25  & \textbf{35.63}  & 0.26 & \textbf{30.22} & 0.50  & \textbf{37.52} & 2.29  & \textbf{37.91} & 1.23  & \textbf{49.67}  & 2.41  &
        \textbf{33.19} & 0.14  & {31.77} & 0.12  & \textbf{24.86} & 1.85  & ${\textbf{37.48}}_{+6.29}$  & $0.85_{-0.01}$ \\
        \specialrule{0em}{0.04mm}{0.04mm}
    \bottomrule
    \end{tabular}%}
    \label{tab:ICS-PosNeg}
    \vspace{-0.25cm}
\end{table*}

\vspace{-0.25cm}
\subsection{Interactive Community Search}
\label{sec.exp.ics}
\vspace{-0.1cm}
\blue{ICS-GNN \cite{gao2021ics_ICS-GNN} is a recent GNN-based model for interactive community search. Given a query, ICS-GNN returns an answer community.  If the user is not satisfied with the answer, he/she can give a feedback (e.g., adding some additional vertices), and then ICS-GNN will respond with a revised answer.  This interaction continues until the user is satisfied.  In each interaction, ICS-GNN first finds a candidate subgraph, learns the vertex embedding through a  Vanilla GCN model \cite{Kipf_GCN} and finally employs a BFS based algorithm to select $k$-sized community with the maximum GNN scores. 
Note that ICS-GNN does not use any training queries with ground-truth communities to train the model; for each user query, it re-trains the GNN model to obtain the vertices' embeddings only from the knowledge of the given query.  } ICS-GNN only supports non-attributed community search. 

In this experiment, we replace the Vanilla GCN model \cite{Kipf_GCN} in ICS-GNN with our community search models \QD and \AQD to compare the performance of interactive community search.

\vspace{-0.2cm}
\subsubsection{Performance in Effectiveness} We first use \QD to replace the GNN model in ICS-GNN framework for non-attributed community search.  As shown in Table~\ref{tab:ICS-PosNeg}, \QD outperforms the original ICS-GNN in all data sets with $2.37\%$ improvement in F1-score.
We also use \AQD to replace the GNN model in ICS-GNN so it supports interactive attributed community search. As shown in Table~\ref{tab:ICS-PosNeg}, \AQD further improves the F1-score of the original ICS-GNN for all data sets by $3.72\%$ (AFN) and $6.29\%$ (AFC) on average. This experiment proves that our \QD and \AQD models are more effective than Vanilla GCN in the ICS-GNN framework.

\vspace{-0.2cm}
\subsubsection{Performance in Efficiency} 

We report the average time of \blue{community search per interaction} by ICS-GNN, \QD and \AQD in Table~\ref{tab:ICS-PosNeg}. The running time of the three models are very close.  Without increasing the time cost, we improve the performance of ICS-GNN and extend it to support attributed community search.

\vspace{-0.2cm}
\subsection{ACS on Large Graphs} \label{sec.exp.large}
\vspace{-0.1cm}

In this experiment, we evaluate the performance of our model for ACS on large graphs. We design a subgraph training mechanism to train our models on large graphs. We first select neighbors of query vertices as the candidate subgraph for each query. According to the number of neighbors, we select 1 or 2-hop neighbors in the fusion graph described in Section \ref{Sec.ACS.CI}. Then we train our model on these small subgraphs and predict communities. %All the three proposed models can equipped with this sub-graph training mechanism. 

We compare \AQD with \ACQ and \ATC for attributed community search on Reddit and an enlarged version of Reddit, denoted as Enlarged\_Reddit. To enlarge Reddit and preserve the ground-truth communities at the same time, we add some new vertices for edges within a community. A new vertex is linked to the two ends of an edge, and the attributes of the new vertex are the average attribute values of the two ends. 
The Enlarged\_Reddit has 3.12M vertices and 126M edges. 
%Since attributes in Reddit and Enlarged\_Reddit are numerical, while \ACQ and \ATC only support keyword attributes. We perform a cut-off on attributes with a threshold 0.5. 

Table~\ref{tab:Large_graphs} reports the index/training time, query time and F1-score of the discovered communities. 
\ACQ takes only 42.4 seconds and 852.7 seconds to build index on Reddit and Enlarge\_Reddit. But in terms of the query time, it costs 32.2 milliseconds and 5726.6 milliseconds, while \AQD only costs 6.7 milliseconds and 5.3 milliseconds respectively. It is worth noting that \ACQ runs out of memory for 25 out of 100 queries on a 300GB memory server. This is because \ACQ finds a $k$-core community with the largest $k$ containing the query vertices. The $k$-core community can be quite large, for example, for a query vertex, \ACQ first finds a 2-core community with more than 800 thousand candidate vertices. The average F1-score of \ACQ is much lower than that of our method in both data sets. ATC did not finish building its index in 7 days and we treat it as timed out. 
From this experiment, we can see that \AQD achieves a good balance between training time and query time in large graphs, and its F1-score is the best.

\renewcommand\arraystretch{0.98}
% \vspace{-0.35cm}
\begin{table}[t]
\footnotesize
    \centering
    \caption{\blue{The performances of ACS methods on large data sets.}}%of different CS and ACS methods
    \vspace{-0.43cm}
    \setlength{\tabcolsep}{0.99mm}
    \blue{
    \begin{threeparttable}
    \begin{tabular}{c||c|c|c||c|c|c}
    \toprule
        &   \multicolumn{3}{c||}{ Reddit}  & \multicolumn{3}{c}{ Enlarged\_Reddit}  \\
        \cline{2-7}
        { Methods} & Index/Train & Query & F1- & Index/Train & Query &  F1-   \\
        &  Time & Time & score & Time& Time & score  \\
        \hline
        % \midrule
        \ACQ & \textbf{42.4 s} & 32.2 ms  & 0.53 & \textbf{852.7 s} & 5726.6 ms*  & 0.38 \\
        % \hline
        \ATC\tnote{\#} & -& -& -& -& -& -  \\
          \AQD  & 4993.6 s & \textbf{6.7 ms} & \textbf{0.91} & 3898.5 s & \textbf{5.3 ms} & \textbf{0.91}\\ %
        %  & &  \\ 
        \specialrule{0em}{0.0mm}{0.0mm} 
    \bottomrule
    \end{tabular}%}
    \vspace{-0.03cm}
    \begin{tablenotes}
    \footnotesize{
    \item[*] 25 out of 100 queries are out of memory when processing. The query time are the average of the rest 75 queries. 
    \vspace{-0.05cm}
    \item[\#] ATC did not finish building its index in 7 days on both two data sets.}
    \end{tablenotes}
    \end{threeparttable} 
    }
    \label{tab:Large_graphs}
    \vspace{-0.5cm}
\end{table}

\vspace{-0.25cm}
\subsection{Ablation Study} \label{sec.ablation}
\vspace{-0.1cm}
In this section, we report the ablation studies of our models, including the effectiveness of Feature Fusion, the sensitivity test of the parameter $\gamma$, \blue{the data split ratio, the epoch number and dropout rate} in the attributed community search task. 

\begin{figure}[t]
\vspace{-0.15CM}
	\centering
	%  % \hspace{-0.04\textwidth}
	\begin{subfigure}[b]{0.49\linewidth}
		\centering
% 		\hspace{-0.05cm}
% 		\includegraphics[width=1.0\textwidth]{./figure/Abla/ablat_f1_2.pdf}
		\includegraphics[width=1.0\textwidth]{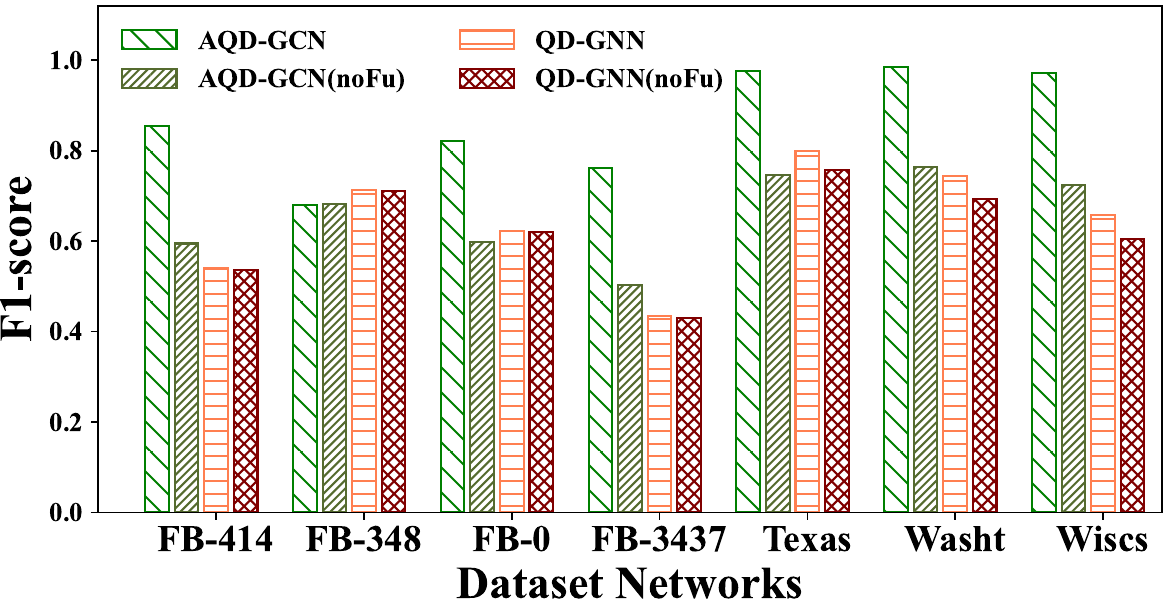}
		\vspace{-0.55CM}
		\caption{F1-score w/wo Feature Fusion.}
		\label{fig:ablation}
	\end{subfigure}
	\begin{subfigure}[b]{0.49\linewidth}
		\centering
% 		\hspace{0.05cm}
% 		\includegraphics[width=1.0\textwidth]{./figure/Abla/threshold.pdf}
		\includegraphics[width=1.0\textwidth]{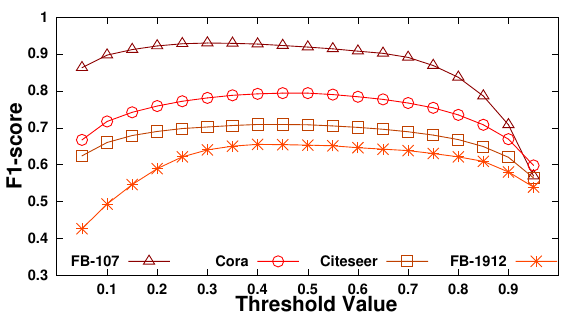}
	       \vspace{-0.55CM}
		\caption{F1-score by varying $\gamma$.}
		\label{fig:threshold}
	\end{subfigure}
	\vspace{-0.45CM}
	\caption{Ablation studies for Feature Fusion and $\gamma$.}
	\label{fig:framework}
	\vspace{-0.35cm}
\end{figure}
%
%the setting of threshold $\gamma$, \blue{the sensitivity of data split ratio, the epoch number and dropout rate}. 
%The results are all under the attributed community search setting.
%
\vspace{-0.25cm}
%\textbf{Initialize} distance compare with just {0,1}
\subsubsection{Ablation Study for Feature Fusion}
% To verify the necessity of fusion operator in \QD and \AQD, we defer more discussions and report results in Figure \ref{fig:ablation}.
In our model, we use the aggregation result $\boldsymbol{h}_{FF}$ in Eq.~\eqref{eq:FuseCS-GCN} in \QD and Eq.~\eqref{eq:ACS-GCN-Fuse} in \AQD to fuse all information, and assign fused features to Query Encoder and Attribute Encoder in Eq.~\eqref{eq:hsl} and Eq.~\eqref{eq:fuse_AE}. To verify the effectiveness of Feature Fusion, we compare the original \AQD and \QD models with \AQD-noFu and \QD-noFu where the encoders do not aggregate in the hidden layer. They only aggregate after the last layer to output the final results. 
% ${\boldsymbol{h}_Q}^{(l)}$ and ${\boldsymbol{h}_N}^{(l)}$ are just ${\textbf{H}_Q}^{(l)}$ and ${\textbf{H}_V}^{(l+1)}$ in Eq.~\eqref{eq:hsl} and \eqref{eq:hvl}. 

The comparison results are shown in Figure \ref{fig:ablation}. For the non-attributed community search problem, the effect of Feature Fusion is more significant in citation networks than that in Facebook ego networks. 
In \QD, Feature Fusion aims to mix the global graph information and local query knowledge. The Facebook ego networks themselves are local graphs. Therefore, Feature Fusion in \QD can only improve the model slightly on Facebook ego networks.
For ACS problem, \AQD outperforms \AQD-noFu substantially in both Facebook ego networks and citation networks. This is because Feature Fusion in \AQD not only fuses the global graph feature, local query structure and similar attribute information at the same time, but also processes query vertices and query attributes simultaneously through the updating of Query Encoder and Attribute Encoder by fused features. This fusion and updating operations significantly improve the results.

% \vspace{-0.15cm}
% \subsubsection{Ablation Study for Aggregation Function}
% In Feature Fusion operator, there is an aggregation function $\AGG(\cdot)$ to aggregate different features in Eq.~\eqref{eq:FuseCS-GCN} of \QD and Eq.~\eqref{eq:ACS-GCN-Fuse} of \AQD. There are many choices for the aggregation operation, such as $\mathrm{Concatenate}$, $\mathrm{Add}$, $\mathrm{Max}$, $\mathrm{Min}$, $\mathrm{Mean}$.  
% %We can choose concatenate, add, max, min or average as the operation of aggregation. 
% The comparison of different aggregation function in graphs is shown in Figure \ref{fig:aggre}. 
% % We adopt attribute query setting in Section\ref{sec.exp}, which generates attribute queries from query vertices. 
% As shown in Figure \ref{fig:aggre}, the $\mathrm{Concatenate}$ performs best in these graphs, which retains all information from three encoders. The $\mathrm{Max}$ function always gains the lowest F1-score, since the output of Graph Encoder in the first layer is always the largest and the $\mathrm{Max}$ function will lose query information.

\vspace{-0.1cm}
\subsubsection{Ablation Study for the threshold $\gamma$}
When translating the model output vector $h_q$ from $\mathbb{R}^n$ to the community vertex set $\hat{\mathcal{C}}_q$ in Section~\ref{sec.frame.onlineQuery}, we use a threshold \blue{$\gamma \in [0,1]$} in the constrained BFS: if $h_{qi}\geq \gamma$, then vertex $\mathrm{v}_i \in \hat{\mathcal{C}}$, otherwise $\mathrm{v}_i \notin \hat{\mathcal{C}}$. In the experiments, we choose $\gamma$ which achieves the best performance in the validation set. To analyze the impact of the threshold, we vary $\gamma$ from 0.05 to 0.95 and report the F1-score in Figure \ref{fig:threshold}. When $\gamma$ is between 0.3 and 0.7, there is very little fluctuation in the performance. Therefore, \AQD is not very sensitive to the setting of this threshold.

\begin{figure}[t]
\vspace{-0.25cm}
\small
	\centering
	\begin{subfigure}[b]{0.49\linewidth}
		\centering
		\includegraphics[width=1\textwidth]{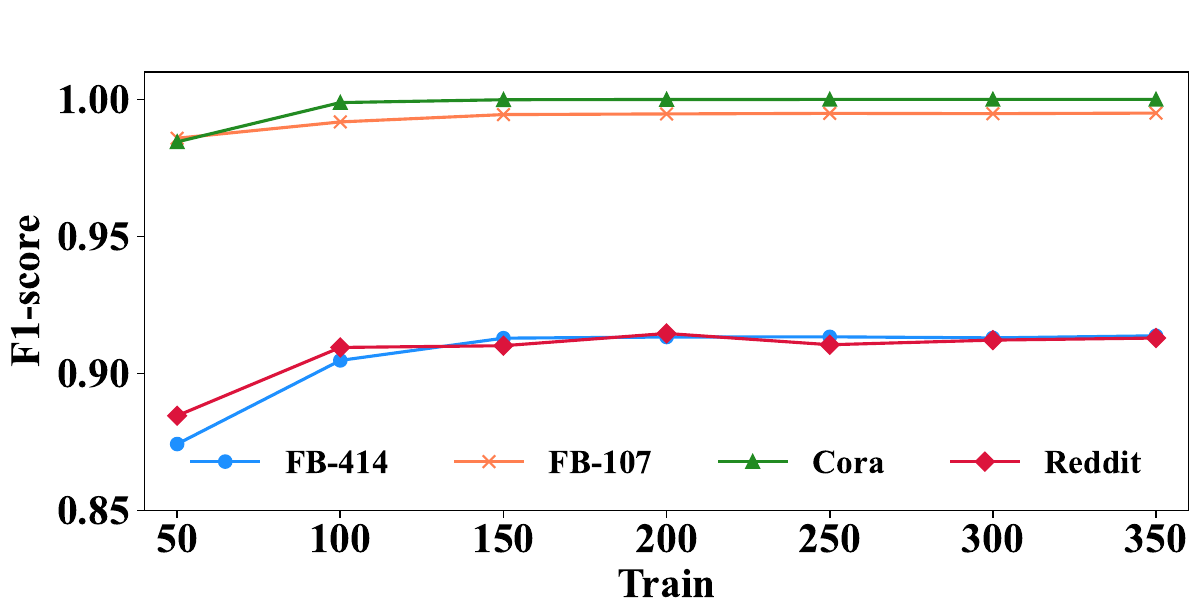}
		\vspace{-0.55cm}
		\caption{\blue{Vary training set size.}}
		\label{fig:train25-150}
	\end{subfigure}
	\begin{subfigure}[b]{0.49\linewidth}
		\centering
% 		\hspace{-0.1cm}
% 		\includegraphics[width=1.\textwidth]{./Fig/val25-100.pdf}
% 		\includegraphics[width=1\textwidth]{./figure/valid.pdf}
		\includegraphics[width=1\textwidth]{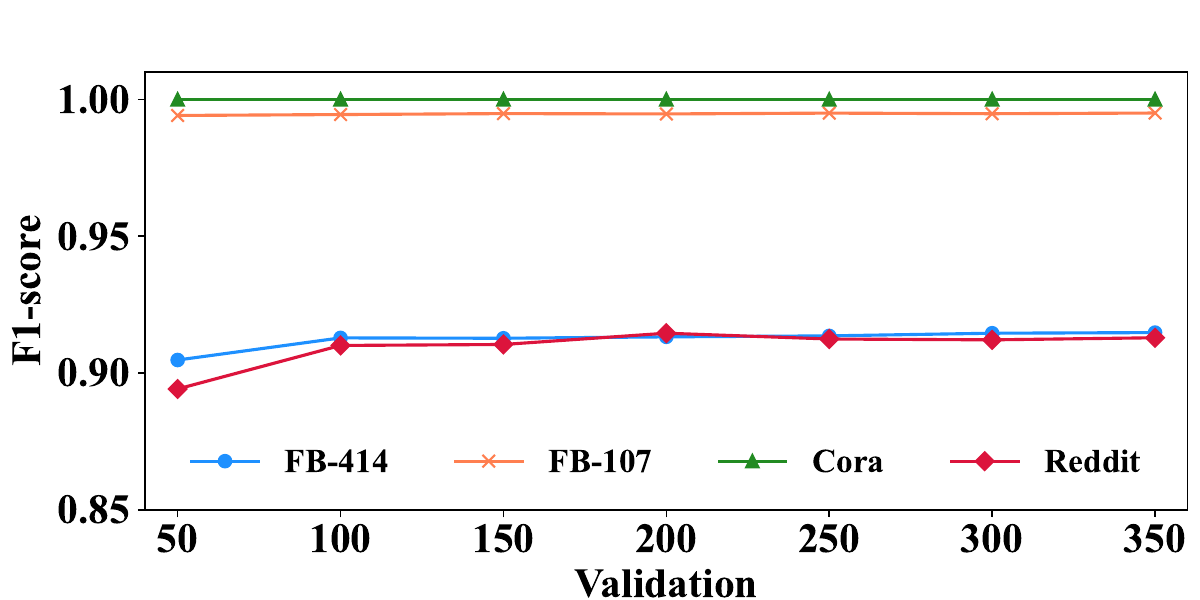}
		\vspace{-0.55cm}
		\caption{\blue{Vary validation set size.}}
		\label{fig:val25-100}
	\end{subfigure}
	\vspace{-0.4cm}
	\caption{\blue{Ablation study for data split ratio.}}
	\label{fig:train:val:test}
	\vspace{-0.25cm}
\end{figure}

\vspace{-0.15cm}
\blue{
\subsubsection{Ablation Study for the data split ratio}
In all the experiments above, we fix the training/validation/test size ratio as 150:100:100. To test the sensitiveness of data split ratio, we vary the training set size from 50 to 350, and fix both the validation and test set size as 100. The results are plotted in Figure \ref{fig:train25-150}. We also vary the validation set size and fix the training set size as 150 and the test set size as 100. The results are plotted in Figure \ref{fig:val25-100}.

When the training set size increases from 50 to 100, the F1-score of all data sets has a notable increase, but when the training set size further increases from 100 to 350, the F1-score remains quite stable.  When varying the validation set size, for Cora and FB-107 the F1-score remains stable; for FB-414 and Reddit, the F1-score increases when the validation set size increases from 50 to 100, and then remains stable afterwards.

This experiment shows that when the training/validation set is very small, increasing the size can improve the performance; but when the size is above 100, the performance remains stable.
%Overall, with the growth of the training/validation data set, the performances are improved. And our model performs quite stable with different ratios of train/validation/test splits. 
}

\begin{figure}
% \vspace{-0.15cm}
	\centering
	\begin{subfigure}[b]{0.49\linewidth}
		\centering
		\includegraphics[width=1.\textwidth]{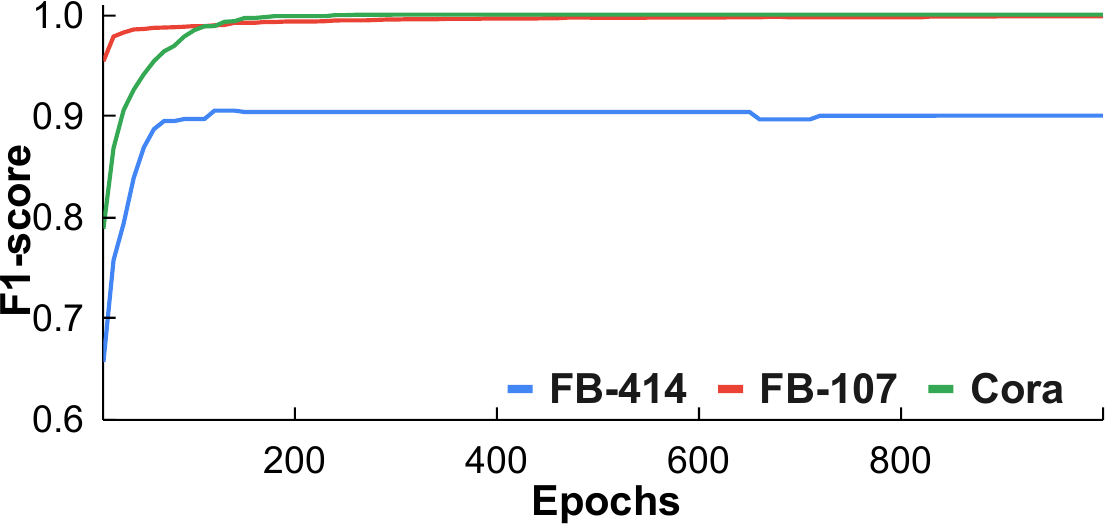}
		\vspace{-0.55cm}
		\caption{\blue{Vary the epoch number.}}
		\label{fig:epoch}
	\end{subfigure}
	\begin{subfigure}[b]{0.49\linewidth}
		\centering
% 		\hspace{-0.1cm}
		\includegraphics[width=1.\textwidth]{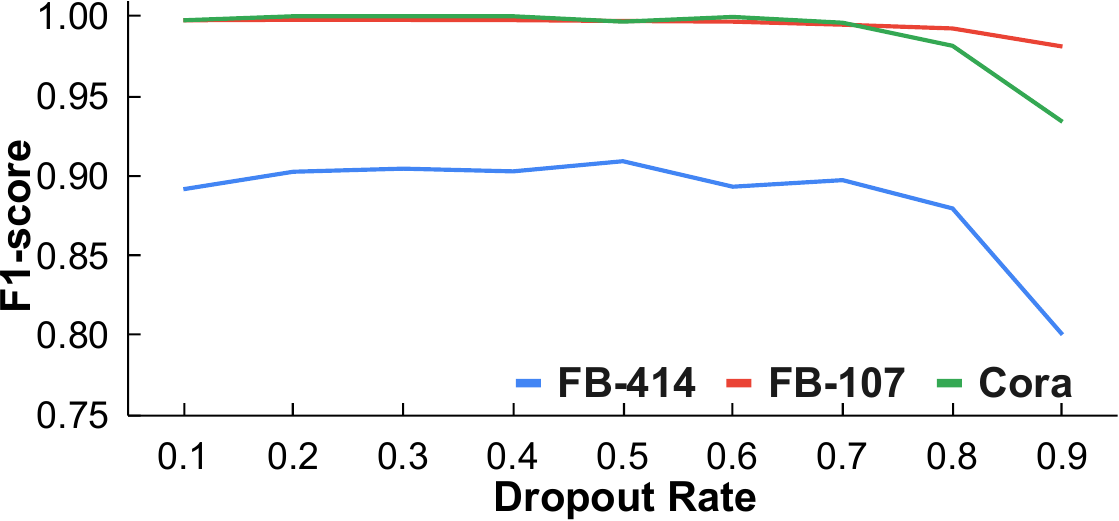}
		\vspace{-0.55cm}
		\caption{\blue{Vary the dropout rate.}}
		\label{fig:dropout}
	\end{subfigure}
	\vspace{-0.4cm}
	\caption{\blue{Ablation studies for epoch number / dropout rate.}}
	\label{fig:epoch:dropout}
	\vspace{-0.45cm}
\end{figure}

\vspace{-0.15cm}
%\textbf{Initialize} distance compare with just {0,1}
\blue{
\subsubsection{Ablation study for the epoch number}
In the model training stage, we set the epoch number as 300. To test the performance with different epoch number, we vary it from 0 to 1000 and show the average F1-score of the test queries in Figure~\ref{fig:epoch}. \AQD achieves very good performance within 100 epochs and converges after 200 epochs. 
% Thus we choose 300 epochs as our setting. 

\vspace{-0.15cm}
\subsubsection{Ablation study for the dropout rate}
For the dropout rate, we vary it from 0.1 to 0.9 and show the F1-score in the test set in Figure~\ref{fig:dropout}. The performance is stable when the dropout rate is from 0.1 to 0.7. If the dropout rate is too larger than 0.7, the model drops too much information and the performance decreases accordingly. Following the general setting of dropout rate \cite{srivastava2014dropout}, we choose the middle value, 0.5, as the default value in our experiments.
}

\vspace{-0.2cm}
\section{Conclusions}\label{sec.con}
\vspace{-0.1cm}
In this paper, we propose the \QD and \AQD for non-attributed community search and attributed community search respectively. In \QD, we first propose a query-driven component to acquire queries directly and avoid the re-training process in the existing GNN-based community search model ICS-GNN. Then we combine the local query-dependent structure and global query-independent vertex embedding. 
For attributed community search, we model vertex attributes as a bipartite graph and further propose the \AQD model. 
% \AQD model can also serve as the GNN model in ICS-GNN and extend ICS-GNN to the interactive attributed community search problem. 
To the best of our knowledge, \AQD is the first GNN model for attributed community search. 
Moreover, we apply \QD and \AQD in the framework of ICS-GNN for interactive attributed community search. 
Experiments demonstrate that the proposed models outperform previous approaches significantly. 
\blue{The proposed models are trained through historical queries (training queries), then applied for online query.  In the future, more research on training query selection can be carried out to train the model with limited training queries for large graphs.  In addition, as time goes by, more historical queries can be collected and the model 
can be updated with them as training queries to improve its performance.  The model update mechanism is worth further study. }

% As a learning based model, \AQD provides a promising solution for attributed community search.
\vspace{-0.5cm}
\begin{acks}
\vspace{-0.1cm}
%  This work was supported by the [...] Research Fund of [...] (Number [...]). Additional funding was provided by [...] and [...]. We also thank [...] for contributing [...].
 The work was supported by grants from NSFC Grant No. U1936205, the Research Grant Council of the Hong Kong Special Administrative Region, China [Project No.: CUHK 14205618], Tencent AI Lab RhinoBird Focused Research Program GF202101, and CUHK Direct Grant No. 4055159. 
 Additional funding was provided by the HK RGC Grant Nos. 22200320 and 12200021. Hong Cheng is the corresponding author.
\end{acks}

%\clearpage

\bibliographystyle{ACM-Reference-Format}
\bibliography{sample}

\end{document}